\documentclass[showpacs, aps, prb, unsortedaddress,showkeys,longbibliography]{revtex4-2}

\usepackage{amssymb}
\usepackage{latexsym}
\usepackage{amsmath}
\usepackage{graphicx}
\usepackage{subfigure}
\usepackage{float}
\usepackage{color,soul}
\usepackage{amsthm}
\usepackage[dvipsnames]{xcolor}

\newtheoremstyle{named}{}{}{\itshape}{}{\bfseries}{.}{.5em}{\thmnote{#3's }#1}
\theoremstyle{named}

\begin{document}
%\tableofcontents
\newpage
\title{The Analytical Structure of Acoustic and Elastic Material Properties}

\date{\today}
\author{Hossein Khodavirdi}
\author{Ankit Srivastava}
\thanks{Corresponding Author}
\affiliation{Department of Mechanical, Materials, and Aerospace Engineering,
Illinois Institute of Technology, Chicago, IL, 60616
USA}
\email{asriva13@iit.edu}

\begin{abstract}
In this paper, we take an in-depth look at the analytical structure of the material transfer functions which govern acoustic and elastic response. These include wavenumber ($\kappa$) in such media and refractive index ($n$), density ($\boldsymbol{\rho}$) and its inverse, stiffness ($\boldsymbol{C}$) and compliance ($\boldsymbol{D}$) tensors as well as the Bulk modulus ($B$), and finally the broader generalization of these properties which is now known as the Willis tensor ($\boldsymbol{L}$). Our goal is to clarify the appropriate dispersion relations applicable to these properties from the perspective of passivity. Under some mild assumptions, causality ensures that these properties are analytical in the upper half but deriving dispersion relations for them requires one to know how they behave in the limit $|\omega|\rightarrow\infty$. Unlike electromagnetism, such a determination cannot be made on physical grounds since in that limit the continuum approximation breaks down. Instead, we can exploit the properties of the Herglotz-Nevanlinna (H-N) functions along with their tensorial counterparts which characterize the transfer functions of certain passive systems and for which the appropriate dispersion relation is known. Our aim, therefore, is to clarify the relationship that these transfer functions have with Herglotz functions, which in turn determines the appropriate dispersion relation for them. Our analysis shows that based upon passivity alone, dispersion relations of \emph{minimum} order 1 apply to the Fourier transforms of $\boldsymbol{D},\boldsymbol{\rho}, n'$, and the inverse of $B$, order 3 apply to $\boldsymbol{C},B$, and the inverse of $\boldsymbol{\rho}$, and order 2 applies to $\kappa$.
%, {\color{red}order 0 applies to $n^2$}
\end{abstract}

% \keywords{causality,passivity,metamaterials}

\maketitle

\section{Introduction}\label{sec:introduction}
If a cause-effect relation adopts a convolution form, then the assumption that the effect cannot exist before its cause -- the colloquial statement of causality -- has strong implications for the transfer function of the relationship. Such transfer functions are ubiquitous in physics and engineering. For the purpose of the current study, we will be concerned with the transfer functions which represent dynamic acoustic and elastic material responses. These transfer functions include the compliance, $D$, and stiffness, $C$, of solid materials, bulk modulus, $B$, of a liquid or air, density, $\rho$, of materials, and finally, wavenumber ($\kappa$) and refractive index ($n'$) -- quantities which characterize wave propagation in such materials. They also include the vectorial or tensorial forms of these quantities as well as the general Willis tensor, $\boldsymbol{L}$ \cite{willis2009exact,Srivastava2015CausalityElastodynamics}, which has come to characterize metamaterial response. Considering any one of these quantities as a time dependent function, $m(t)$, causality states that $m(t)=0\forall t<0$. Under some conditions, causality can give rise to relations between the real and imaginary parts of the Fourier transform of $m(t)$. Denoting by $\tilde{m}(\omega)$, the Fourier transform of $m(t)$, causality can result in the following relations~\cite{srivastava2020causality}:
\begin{eqnarray}
\label{eKramersKronigGen}
\Re \tilde{m}(\omega)=\frac{\omega^n}{\pi}\mathcal{P}\int_{-\infty}^{\infty}\frac{\Im \tilde{m}({\omega^{'}})}{\omega^{'n}}\frac{d\omega'}{\omega'-\omega};\quad
\Im \tilde{m}(\omega)=-\frac{\omega^n}{\pi}\mathcal{P}\int_{-\infty}^{\infty}\frac{\Re \tilde{m}({\omega^{'}})}{\omega^{'n}}\frac{d\omega'}{\omega'-\omega}
\end{eqnarray}
In the above, $\mathcal{P}$ is the Cauchy Principal value and the relations are called the generalized Kramers Kronig relationships~\cite{beltrami1966distributionalb,waters2000application}. Here, $n$ is any integer greater than or equal to some integer $l$ and the value of $l$ depends upon the behavior of $\tilde{m}(\omega)$ in the limit $|\omega|\rightarrow\infty$. The main aim of this paper is to establish the correct value of $l$ for the transfer functions alluded to above. A certain value of $l$ has been used in a recent paper \cite{muhlestein2016reciprocity} in the context of Willis tensors but without the supporting arguments necessary. 
% {\color{blue}What I understand from this sentence is that Haberman found a right value for $\boldsymbol{L}$ and the only problem is that they have not talked about supporting arguments, right? If this sentence is right so one can interpret that our paper is wrong because our results are not identical.}

For electromagnetic waves, the situation is markedly simpler due to the limiting speed of light -- a benefit which does not exist in acoustics or elastodynamics~\cite{norris2018integral}. For electromagnetics, the refractive index, $n'(\omega)$, is proportionally related to $\sqrt{\epsilon(\omega)}$ (assuming that the magnetic permeability, $\mu$, is equal to unity). $\epsilon(\omega)$ is in turn related to the susceptibility of the medium $\chi(t)$, which relates the physical quantities, electric polarization and electric field, through a convolution relation. $\chi(t)$ is automatically causal from physical considerations and, therefore, $n'(\omega),\kappa(\omega)/\omega$ are causal as well. Furthermore, since the dielectric is underlined by a vacuum and the high frequency behavior of a wave approaches that of vacuum propagation \citep{nussenzveig1972causality,weaver1981dispersion}, physics dictates that in the high frequency limit, $\epsilon(\omega)$ tends to $1$ (since $\chi(\omega)$ goes down as $1/\omega^2$ and $\epsilon=1+4\pi\chi$). Since $\epsilon(\omega)$ tends to 1 in the high frequency limit, so does $n'(\omega)$. For electromagnetic wave propagation, $n'(\omega)=n_r+i(c_0\beta/2\omega)$, where the real part of $n'$, $n_r$, is called the real refractive index, and the factor $\beta$ is called the extinction coefficient which governs the attenuation of the medium. $c_0$ is the speed of light which is a constant. In the high frequency limit, since $n'(\omega)\rightarrow 1$, we have $n'(\omega)-1$ tending to zero. Hence, we see that for electromagnetics, there are fundamental limits on the behavior of material properties such as $\epsilon(\omega)$ and wave properties such as $n'(\omega)$ in the limit $|\omega|\rightarrow\infty$. Determining how these quantities behave in the high frequency limit allows us to determine the correct value of $l$ for these properties.

The same arguments cannot be made for acoustics or elastodynamics. Unlike dielectrics which behave as vacuum in the high frequency limit, acoustic or elastic mediums are not underlined by a reference response in such limits. In fact, it does not make sense to talk about material properties such as compliance or stiffness in such limits because the continuum approximation breaks down. To navigate this problem, researchers in acoustics and elastodynamics have resorted to empirical arguments. For instance, Ginzberg~\cite{ginzberg1955} essentially assumed that $\kappa(\omega)/\omega$ exists as $|\omega|\rightarrow\infty$, and that it approaches some limiting value independent of $\mathrm{arg}\;\omega$, which allowed him to derive a value of $l$ for $\kappa(\omega)$. The same approach was followed by Futterman~\cite{futterman1962dispersive} in his application of dispersion relation to seismic wave propagation (see also~\cite{lamb1962attenuation,strick1967determination,azimi1968impulse,randall1976attenuative,liu1976velocity} for further discussions on dispersion in seismic waves and connections to Kramers-Kronig relationships). His essential argument is that it is difficult to envision that the structure of the Earth would resonate to a disturbance at infinite frequency. This allows him to say that the imaginary part of $n'(\omega)$, which is proportional to attenuation, must be 0 in that limit and the real part must equal some constant $n_r(\infty)$. For acoustic wave propagation, the derivation of the  dispersion relations is often based upon assuming a functional form for attenuation~\cite{hamilton1970sound,horton1974dispersion,horton1981comment}. Consider $\kappa(\omega)=\omega/c(\omega)+i\alpha(\omega)$, where $c(\omega)$ is the phase velocity of the wave, and $\alpha(\omega)$ is the attenuation constant. For media in which the attenuation satisfies a frequency power law, $\alpha(\omega)=\alpha_0|\omega|^y$, the correct value of $l$ depends upon the power coefficient $y$~\cite{waters1999kramers,waters2000application,waters2003differential,waters2005causality}. Thus we see that the determination of $l$ in acoustics and elastodynamics is generally either based upon empirical assumptions on the functional form of the property under consideration or on the high frequency behavior of these properties -- the latter especially being based on a set of assumptions which may be difficult to defend on physical grounds.

There is one especially notable work which attempts to deduce the correct value of $l$ for passive acoustic and elastodynamic media through the use of Herglotz functions~\cite{herglotz1911uber}, sometimes also called Nevalninna or Pick functions. These are functions which are analytic in the upper half of the complex plane where they have non-negative imaginary part. Herglotz functions have well known behavior in the high frequency limit and the corresponding correct value of $l$ which may be derived for them is well known~\cite{youla1958representation,nussenzveig1972causality,beltrami1967linear}. Weaver and Pao~\cite{weaver1981dispersion} showed that for passive media, $\kappa(\omega)$ is a Herglotz function and derived dispersion relations for it. Our work picks up from Weaver and Pao's work and fills in some missing details. Weaver and Pao restricted their discussions to $\kappa(\omega)$ and did not consider other properties of interest described above. This is understandable in part because the ideas of frequency dependent density, for example, have only become popular with the advent of metamaterials~\cite{srivastava2015elastic}. Here, we extend the Herglotz function based analysis that Weaver and Pao pursued to the other properties which control the dynamic behavior of acoustic and elastodynamic media including the Willis tensor. Furthermore, instead of following the Cauchy integral based proofs presented by Weaver and Pao, we present proofs which are based upon distribution theory. These are more succinct and applicable to generalized functions as well. Furthermore, our analysis applies to tensorial properties which were not considered by Weaver and Pao. Our approach in the rest of the paper will be the following: for the properties under consideration, we will show that they are related to Herglotz functions and then we will use the theory of Herglotz functions to derive the correct value of $l$ for these properties thus determining the correct dispersion relations for them. To show that these properties are related to Herglotz functions, we will extensively invoke the principle of passivity. In other words, the main conclusions in this paper only apply to materials and mediums which do not have sources of energy.

\section{Generalized Kramers-Kronig relationships}\label{sec:distributions}

\begin{figure}[!htb]
   \begin{minipage}{0.45\textwidth}
     \centering
     \includegraphics[width=1\linewidth]{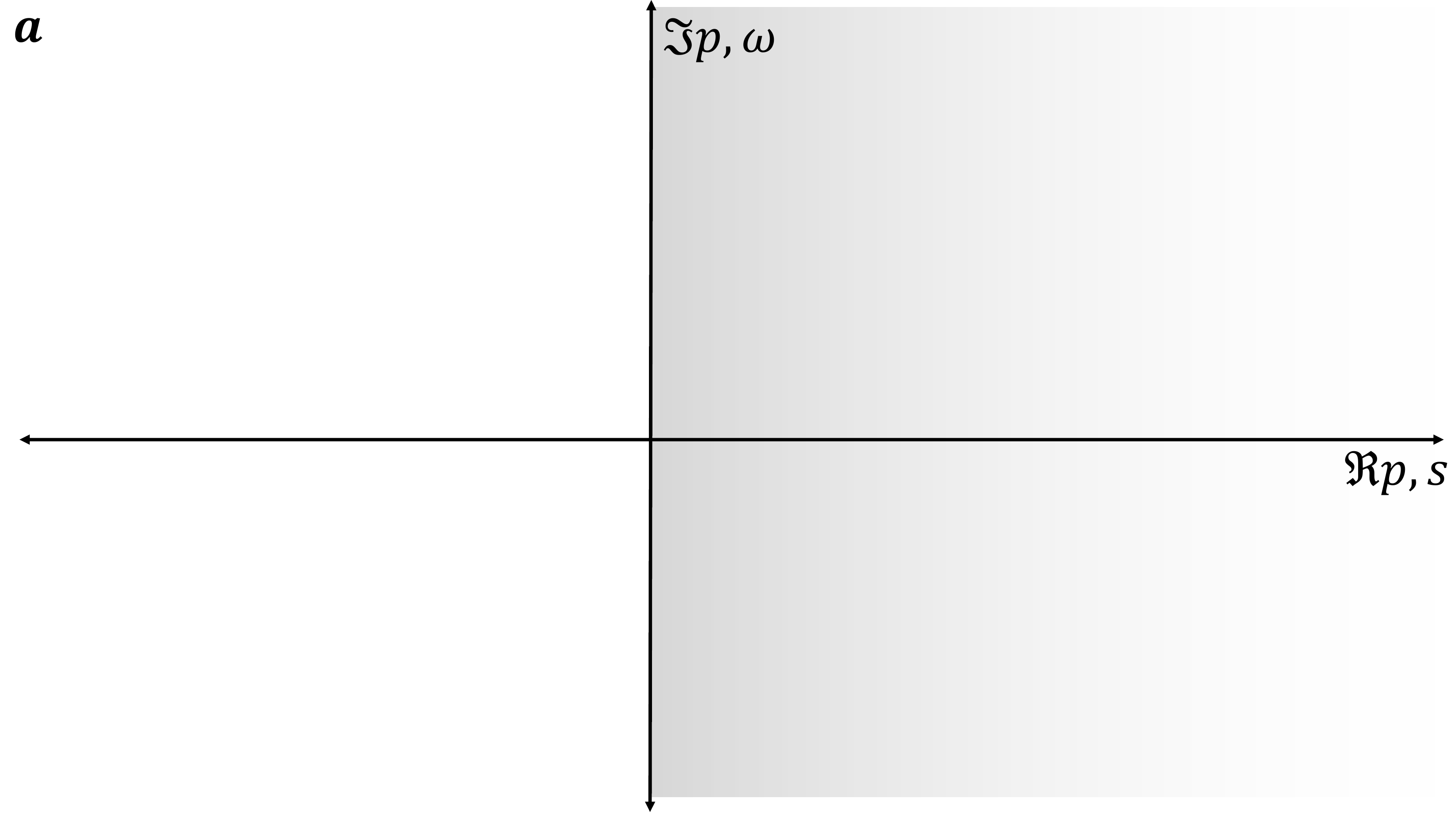}
     
   \end{minipage}\hfill
   \begin{minipage}{0.45\textwidth}
     \centering
     \includegraphics[width=1\linewidth]{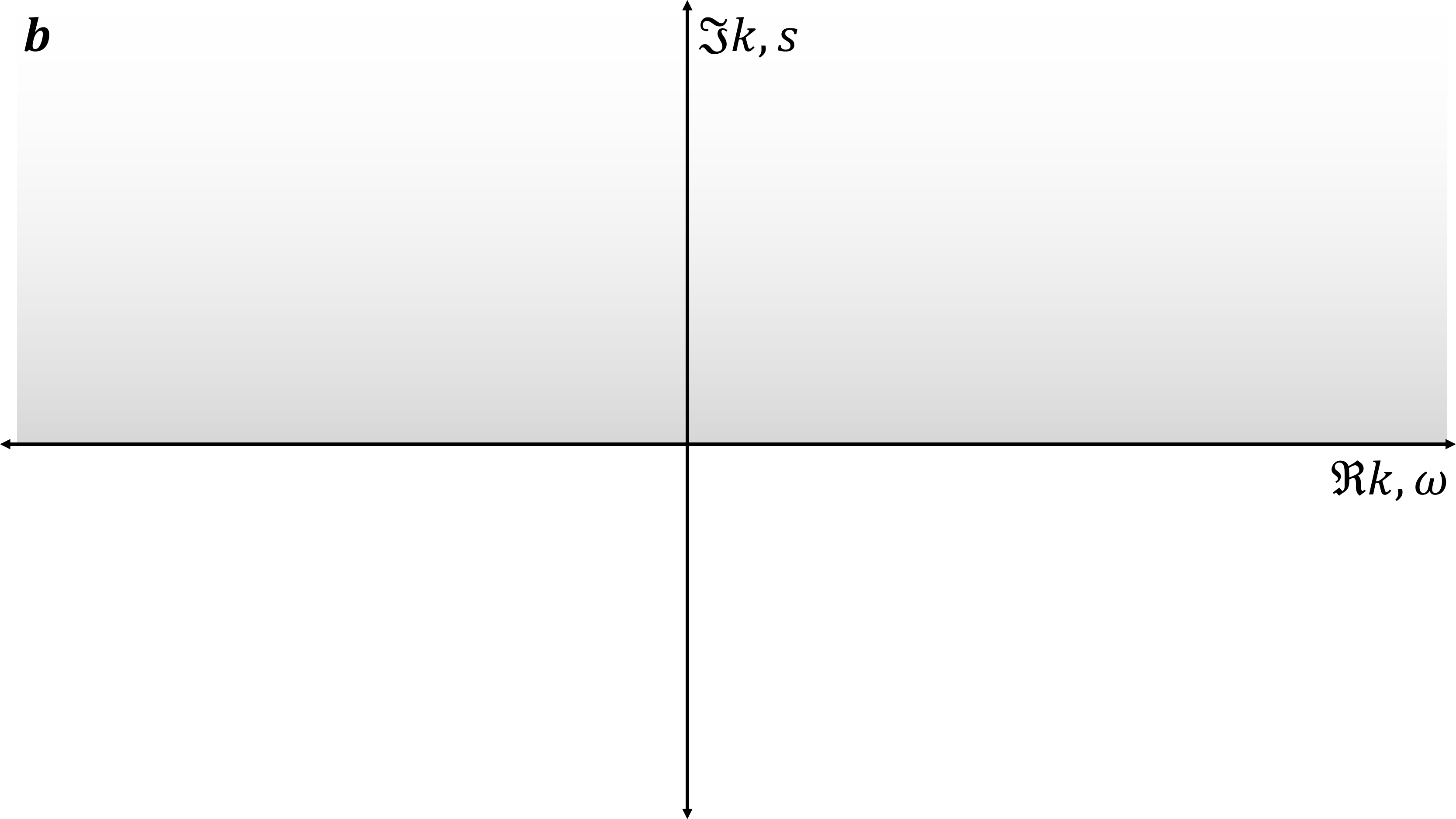}
     %\caption{Interpolation for Data 2}\label{Fig:Data2}
   \end{minipage}
   \caption{Two representation of the complex frequency plane (a)$p=s+i\omega$ and (b)$k=\omega+is$.}\label{Fig1}
\end{figure}

Here we are concerned with the space of temperate distributions and we summarize some relevant results for the same. For more details on distribution theory and the space of temperate distributions, we refer the reader to exhaustive references on the topic~\cite{zemanian1965distribution}. In this section, we only present the immediately useful results without providing any proofs. We define $\mathcal{S}$ as the space of rapidly decreasing test functions characterized by $\phi(t)\in C^\infty$ which, together with all their derivatives, decrease faster than any inverse power of $t$ as $\vert t\vert\rightarrow \infty$:
\begin{eqnarray}
\lim_{\vert t\vert\rightarrow\infty}\vert t^p\frac{\partial^m \phi(t)}{\partial t^m}\vert=0;\quad p,m=0,1,...
\end{eqnarray}
We define the class of temperate distributions $\mathcal{S}'$ as the set of distributions which are linear functionals on $\mathcal{S}$ and the Fourier transform of a distribution $g\in\mathcal{S}'$ through the relation $(\mathcal{F}g,\phi)=(g,\mathcal{F}\phi)$. $(x,y)$ represents the inner-product $\int_{-\infty}^{\infty}xydt$ and $\mathcal{F}\phi$ represents the usual Fourier transform given by:
\begin{eqnarray}
\label{eFourierPoint}
\mathcal{F}\phi(t)=\tilde{\phi}(\omega)=\int_{-\infty}^\infty \phi(t)e^{i\omega t}dt
\end{eqnarray}
Fourier transform of a distribution $g(t)\in\mathcal{S}^{'}$, denoted by $G(\omega),\mathcal{F}g,\tilde{g}$, exists and belongs to $\mathcal{S}^{'}$. With $k=\omega+is$, the Laplace transform is defined through the Fourier transform using $\mathcal{L}g=G(k)=\mathcal{F}(g(t)e^{-st})$. A causal distribution in $\mathcal{S}^{'}$ is one which is zero for $t<0$ and belongs to a subspace of $\mathcal{S}^{'}$ denoted by $\mathcal{S}^{'}_+$. If $g(t)\in\mathcal{S}^{'}_+$ then its Laplace transform has a region of convergence $s>0$ and its boundary value is the Fourier transform $G(\omega)$. Furthermore, the following generalized Hilbert transform applies to $G(\omega)$:
\begin{eqnarray}
\label{eHilbertGen}
G(\omega)=-\frac{\omega^n}{\pi i}\left[\frac{G(\omega)}{\omega^n}*\mathcal{P}\left(\frac{1}{\omega}\right)\right].
\end{eqnarray}
$*$ denotes the convolution operation, $\mathcal{P}$ denotes the principal value distribution and $n$ is any integer greater than or equal to a specific non-negative integer $l$. Equating the real and imaginary parts of (\ref{eHilbertGen}), we arrive at the generalized dispersion relationships (generalized Kramers-Kronig relationships) that were also mentioned in (\ref{eKramersKronigGen}). The value of $l$ depends upon the growth properties of $G(\omega)$ and, equivalently, the discontinuity properties of $g(t)$. To be more specific, we note the result that every distribution is a finite order derivative of a continuous function. $l$ is the order of the derivative which connects $g(t)$ to some continuous function. Once $l$ is determined, one can derive a set of valid dispersion relations by taking $n=l$ in (\ref{eHilbertGen}) and separating the real and imaginary parts. Higher order dispersion relations are also valid if one takes $n>l$, however, dispersion relations with $n<l$ are invalid. Our next step is to collate the set of proofs which establish the correct value of $l$ for Herglotz functions. In what follows, we will use letters (either in lower or higher case) such as $x,f,h$ to represent quantities in the time domain and by hat such as $\hat{x},\hat{f},\hat{h}$ to represent their Laplace or Fourier transforms. Sometimes, we will refer to classical results from electrical networks and control theory where the convention is to refer to the region $s>0$ as the right half plane. This convention emerges from the definition of the complex frequency, $p$ which parametrizes the Laplace transform, as $p=s+i\omega$. In this convention, we will refer to the Laplace transform as $\hat{X}(p)$, as an example. At other times we will refer to the convention more common in physics where the region $s>0$ signifies the upper half plane. This emerges from defining the complex frequency, $k=\omega+is$. In this convention, we will refer to the Laplace transform as $\hat{X}(k)$, as an example. Fourier transform ($\hat{X}(\omega)$, for example) will be indicated by using the dependence on the frequency $\omega$.  The mentioned complex frequency domains are illustrated in Fig. (\ref{Fig1}).
%\begin{figure}
%   \centering
%    \subfigure[]{\includegraphics[width=0.5\textwidth]{images/Fig1a.pdf}} 
%    \subfigure[]{\includegraphics[width=0.5\textwidth]{images/Fig1b.pdf}} 
%    \caption{(a) blah (b) blah}
%    \label{fig:foobar}
%\end{figure}

\section{Correct value of $l$ for Herglotz functions}\label{sec:herglotz}

Here, we introduce the Herglotz integral representation\cite{Herglotzthesis,sumrules,herglotzorig} for functions which are holomorphic in the upper half plane of the complex frequency with positive imaginary part and show \cite{zemanian1963n,meixnernetwork} that their inverse Fourier transforms are a second order derivative of continuous functions. There is a related concept in areas such as control theory where the right half of the complex plane is used more commonly. There, functions which are holomorphic in the right half and which posses a non-negative real part there, are termed positive functions~\cite{CharacterZemanian,Necandsuf}. We emphasize that a Herglotz function is just a mapped version of a positive function. In the rest of this section, we show that a Herglotz function represents the transform of a time domain function which is a second order derivative of a continuous function -- in other words, $l=2$ applies to all Herglotz functions.

% The proof begins by looking into the Herglotz function as a distribution which its action on a testing function in distributional sense assists us to find its representation in time domain. The proof in this section in addition to the point that lots of our desired transfer functions (or similar forms of them) in elastodynamics have properties of a Herglotz function along with the theorem which indicates that temperate distributions are finite order derivatives of a continuous function, assist us to conclude that the necessity of being square integrable for writing a dispersion relation is now relaxed. Moreover, based on the same theorem the order of the derivative in our proof is the one we need to write the generalized form of Hilbert transform as a dispersion relation\cite{beltrami2014distributions}.\par

Consider the Herglotz representation of any function which is holomorphic in the upper half plane of $k$ (complex frequency; $k=\omega+is$) and possesses a non-negative imaginary part there\cite{Herglotzthesis,cassier2017bounds}:
\begin{equation}
    \label{Herg3}
    \hat{h}(k)=a+bk+\frac{1}{\pi}\int_{-\infty}^{+\infty}\frac{1+rk}{r-k}d\nu(r)
\end{equation}
The reader is referred to section \ref{sec:appendix} subsection A, for further information about the derivation of this expression. In Eq. (\ref{Herg3}), $a\in \mathbb{R}$, $b\geq0$ and $\nu$ is a positive bounded measure. It should be noted that this representation is equivalent to the Cauer representation for positive functions~\cite{cauer1932poisson}. The above represents the Laplace transform of the time domain function $h(t)$ whose Fourier transform can be found simply by taking the limit $s\rightarrow 0^+$ of $\hat{h}(k)$. This is done by considering the identity
\begin{eqnarray}
\nonumber
\lim_{s\rightarrow 0^+}\frac{1}{\omega+is}=\mathcal{P}(1/\omega)-i\pi\delta(\omega),
\end{eqnarray}
which allows us to write,
\begin{equation}
    \label{Herg5}
     \hat{h}(\omega)=a+b\omega+\frac{1}{\pi}\int_{-\infty}^{+\infty}\left(1+r\omega\right)\left[P\left(\frac{1}{\omega-r}\right)-i\pi\delta(\omega-r)\right]d\nu(r)
\end{equation}
% In order to develop a proof which shows that a Herglotz function is a second order derivative of a continuous function and also to prevent restrictive conditions which are necessary when working in a strong topology, from this point we no longer talk about functions and those should be interpreted as functionals of a specific type\cite{BeltramiE.J.1966}. From the theory of distributions, we pick the space of all rapidly decreasing test functions (i. e. $\mathcal{S}$) and its dual space, means $\mathcal{S}^{'}$, as the spaces which the test function and its functional are members of them. 
$\hat{h}(\omega)$ is understood as a distribution and in distributional analysis applied to passivity, it is customary to assume that $\hat{h}\in\mathcal{S}^{'}$. We can now write the equation above using distributional notation:
\begin{equation}
\begin{split}
\label{Herg6}
    (\hat{h}(\omega),\phi(\omega))=(a,\phi(\omega))+(b,\phi(\omega))-2i\left(\int_{-\infty}^{\infty}(1+r\omega)\left[\frac{1}{2}\delta(\omega-r)-\frac{1}{2\pi i}P\left(\frac{1}{\omega-r}\right)\right]d\nu(r),\phi(\omega)\right)
\end{split}
\end{equation}
where $\phi \in \mathcal{S}$. $(.,.)$ is the inner product, linear with respect to both terms. We'd like to derive the inverse Fourier transform of the above. The inverse Fourier transform of the first term is simply $a\delta(t)$ whereas for the second term, it is $b\delta^{(1)}(t)$ with $\delta^{(1)}(t)$ being the first derivative of $\delta(t)$. To find the inverse Fourier transform of the last term, we recall the following relevant identities for distributional Fourier transforms\cite{nussenzveig1972causality}
\begin{equation}
\label{identity1}
    \left(\mathcal{F}\{f(t)\},\phi(\omega)\right)=\left(f(t),\tilde{\phi}(t)\right);\quad     \mathcal{F}\left\{\frac{e^{-i\xi t}\theta(t)}{2\pi}\right\}=\frac{1}{2}\delta(\omega-\xi)-\frac{1}{2\pi i}P\left(\frac{1}{\omega-\xi}\right)
\end{equation}
This allows us to transform the integral term in Eq. (\ref{Herg6}) as
\begin{equation}
\label{Herg7}
-2i\left(\iint_{-\infty}^{\infty}(1+r\omega)\left[\frac{\theta(t)e^{-irt}}{2\pi}\right]e^{i\omega t}dt d\nu (r),\phi(\omega)\right)
\end{equation}
which can be written in the following form after noting that $\mathcal{F}\{(i\omega)^nf(\omega)\}=\tilde{f}^{(n)}(t)$:
% Now by using the definition of Fourier transform from the Eq. (\ref{eFourierPoint}) one can re-write Eq. (\ref{Herg7}) as,

% \begin{equation}
% \label{Herg8}
% \frac{-i}{\pi}\left[\left(\int_{-\infty}^{\infty}\theta(t)e^{-irt} d\nu (r),\tilde{\phi}(t)\right)+\left(\iint_{-\infty}^{\infty}r\omega\theta(t)e^{-irt}e^{i\omega t}dt  d\nu (r),\phi(\omega)\right)\right]
% \end{equation}
% The second term of Eq. (\ref{Herg8}) also can be simplified more by recalling the identity $\mathfrak{F}\{(i\omega)^nf(\omega)\}=\tilde{f}^{(n)}(t)$. Then we have:

\begin{equation}
\label{Herg9}
\frac{-1}{\pi}\left[i\left(\int_{-\infty}^{\infty}\theta(t)e^{-irt} d\nu (r),\tilde{\phi}(t)\right)+\left(\int_{-\infty}^{\infty}r\theta(t)e^{-irt}  d\nu (r),\tilde{\phi}^{(1)}(t)\right)\right]
\end{equation}

\noindent To proceed further, we recall the classical result that in
\begin{equation}
    \label{FourierStieltje1}
    P(t)=\int_{\infty}^{+\infty}e^{-i\xi t}dF(\xi),
\end{equation}
if the function $F$ is a function with bounded variations on $(-\infty,+\infty)$, then $P(t)$ -- called the after-effect function -- will be bounded and continuous~\cite{meixnernetwork,Necandsuf,CharacterZemanian,phillips1950fourier}.
% In this part and before taking the next step we recall that the aim of this process is to find the time domain representation of our Herglotz function and show that it is a finite order derivative of a continuous function. Here, we take advantage of the method by which Meixner\cite{meixnernetwork} and Zemanian\cite{Necandsuf,CharacterZemanian} showed that a positive function or matrix in its time domain representation is a finite order derivative of a continuous function, called after-effect function in reference\cite{meixnernetwork}, which itself is represented by a Fourier-Stieltjes integral\cite{phillips1950fourier} like:   
% \begin{equation}
%     \label{FourierStieltje1}
%     P(t)=\int_{\infty}^{+\infty}e^{-i\xi t}dF(\xi)
% \end{equation}
% For this integral, if the function $F$ is a function with bounded variations on $(-\infty,+\infty)$ then the after-effect function will be bounded and continuous. Moreover, if $F$ is a non-decreasing function with bounded variations, this time $P(t)$ will be continuous, bounded and positive definite. Using the continuity of $P(t)$ in Eq.(\ref{FourierStieltje1}), we define two more continuous functions like:
% \begin{eqnarray}
% \label{FourierStieltje2}
%     P_1(t)=\int_{-\infty}^{\infty}\cos(\xi t)dF(\xi)\\
%     P_2(t)=\int_{-\infty}^{\infty}\sin(\xi t)dF(\xi)
% \end{eqnarray}
Integration by parts in Eq. (\ref{Herg9}) gives:
\begin{equation}
\label{Herg10}
-i\int_{-\infty}^{\infty}rd\nu(r)\int_{-\infty}^{\infty}\frac{\theta(t)}{r}e^{-irt} \tilde{\phi}^{(2)}(t)dt+i\int_{-\infty}^{\infty}rd\nu(r)\int_{0}^{\infty}\frac{\tilde{\phi}^{(2)}(t)}{r}dt
\end{equation}
Substituting Eq. (\ref{Herg10}) into Eq. (\ref{Herg9}) and then substituting the new form of Eq. (\ref{Herg9}) into Eq. (\ref{Herg6}) and by using identity in Eq. (\ref{identity1}), we can write the inverse Fourier transform of $h$ as:
\begin{equation}
\label{Herg11}
h(t)=a\delta(t)+b\delta^{(1)}(t)-\frac{i}{\pi}\left[P(t)\theta(t)-\frac{d^2}{dt^2}\left(P(t)\theta(t)\right)+\frac{d^2}{dt^2}\left(\theta(t)P(0)\right)\right]
\end{equation}
thus showing that $h(t)$ is a second order derivative of a continuous function, or that $l=2$ applies for $\hat{h}(\omega)$.
% At this point, $h(t)$ is called a distribution of C-order equal to $0$ which means that second order primitives of $h(t)$ are all continuous functions\cite{Necandsuf,BeltramiE.J.1966}.

\subsection{Symmetric Herglotz representation}

For physical applications, $h(t)$ generally represents a real transfer function. If we insist that $h(t)$ is real then we have the following symmetry relations on its Laplace transform:

% One of the most common assumptions in physics and specifically in elastodynamics is to consider the transfer function a function which only takes real values in time domain. (which a transfer function here means any function which relates the input to the output (force to response, or strain to stress etc.)). In other words, considering the values of a stiffness/compliance matrix or the Young's modulus to have only real values in their time representation does make sense. This assumption automatically means:
\begin{equation}\label{real1}
\hat{h}^*(k)=\hat{h}(-k^*)
\end{equation}
where complex conjugation is implied by $*$. Immediately one can write:
\begin{equation}
    \label{real2}
    \hat{h}(k)=\frac{1}{2}\left[\hat{h}(k)+\hat{h}^*(-k^*)\right]
\end{equation}
Now the integral representation of $\hat{h}(k)$ can be modified by applying the above identity to Eq. (\ref{Herg3}):
\begin{equation}
    \label{real4}
    \hat{h}(k)=a+\int_{-\infty}^{+\infty}\frac{1+k^2}{r^2-k^2}rd\nu(r) =     \hat{h}(k)=a+\frac{1}{2}\int_{-\infty}^{+\infty}(1+k^2)\left(\frac{1}{r+k}+\frac{1}{r-k}\right)d\nu(r)
\end{equation}
The above is called a symmetric Herglotz representation\cite{sumrules}.
It can be shown that its inverse transform is given by:
\begin{equation}
    \label{real6}
    h(t)=a\delta(t)-P_2(t)\theta(t)+ \frac{d^2}{dt^2}\left[P_2(t)\theta(t)\right]
\end{equation}
where $P_2(t)=\int_{-\infty}^{\infty}\sin(\xi t)dF(\xi)$. It clear that in the above, $h(t)$ is a fully real function and is, again, a second order derivative of a continuous function. Therefore, by keeping in mind the generalized form of Hilbert pairs in Eq. (\ref{eHilbertGen}) and noting the findings of this section one can write the dispersion relation for any Herglotz function as:
\begin{eqnarray}
\label{dispersion1}
\hat{h}(\omega)=-\frac{\omega^2}{\pi i}\left[\frac{\hat{h}(\omega)}{\omega^2}*\mathcal{P}\left(\frac{1}{\omega}\right)\right].
\end{eqnarray}
% or by separating the real and imaginary parts one can write the Hilbert pairs as\cite{BeltramiE.J.1966}:

% \begin{equation}
% \label{Hilbertpair1}
%  \begin{split}
% \frac{d^{2}}{d\omega^{2}}Img\left[h(\omega)\right]=\frac{1}{\pi}\left(Img\left[h\right]\right)*\left(\frac{d^2}{d\omega^2}\mathcal{P}\left(\frac{1}{\omega} \right)\right) \\
% \frac{d^{2}}{d\omega^{2}}Re\left[h(\omega)\right]=-\frac{1}{\pi}\left(Re\left[h\right]\right)*\left(\frac{d^2}{d\omega^2}\mathcal{P}\left(\frac{1}{\omega} \right)\right) 
%   \end{split}
% \end{equation}

%{\color{red}References were checked upto here.}

%\section{Herglotz functions and passivity}\label{sec:herglotzSI}
\section{Passivity Considerations}\label{sec:herglotzSI}
%{\color{blue} UNDER CONSTRUCTION...}
%{\color{gray}Passivity has two forms. And immittance automatically means Herglotz. Talk in full tensor notation. Is scattering form also related to Herglotz or some related function?}
Transfer functions of passive systems are Herglotz functions under certain considerations, which we discuss in this section.
% In this section, first we discuss two different but equivalent formalism in studying a physical system which each of them end up with two different statements for the same property. Afterwards, the properties of those formalisms along with the definition of positive real distributions in the tensorial form are introduced. This sections ends with a review on the integral representation of positive real matrices on top of a similar representation for positive real matrix distributions which the elements are holomorphic in the upper half-plane. 
Here, we consider all relevant quantities in tensorial and distributional forms. A tensor of distributions $\boldsymbol{f}(t)$ is defined through its actions on a test function $\phi(t)$, both in appropriate spaces. Specifically, $\langle\boldsymbol{f}(t),\phi(t)\rangle$ is the matrix of complex numbers obtained by replacing each element of $\boldsymbol{f}(t)$ by the number that this element assigns to the testing function $\phi(t)$ through the inner product operation. Zemanian introduced tensorial distribution spaces to admit tensors of distributions of appropriate ranks. For example, $\mathcal{S}^{'}_{n\times n\times n\times n}$ is the space of all fourth order tensors whose elements are distributions in $\mathcal{S}^{'}$ etc. Zemanian showed that a single-valued, linear, time-invariant, and continuous input output relation can be written in the convolution form, $\boldsymbol{v}=\boldsymbol{z}*\boldsymbol{j}$, where $\boldsymbol{v,z,j}$ are tensors of distributions in appropriate spaces, and $*$ denotes a convolution in time as well as appropriate tensorial contraction:
\begin{eqnarray}
\label{eMatrixConvolution} 
\boldsymbol{v}=\boldsymbol{z}*\boldsymbol{j}: v_l(t)=z_{lm}(t)*j_m(t);\quad l,m=1,2...n
\end{eqnarray}
%  Particularly, a vector valued distribution, $\boldsymbol{u}$, is a member of $\mathcal{D}^{'}\otimes C^n$ which is a linear and continuous mapping on test functions of $\mathcal{D}$ to the complex n-space. Accordingly, a matrix valued distribution $\boldsymbol{m}$ is a member of $\mathcal{D}^{'}\otimes \mathcal{L}$ which $\mathcal{L}$ is the space of $n\times n$ matrices in either $\mathcal{L}(C^n,C^n)$ form or $\mathcal{L}(R^n,R^n)$. Moreover, writing $\boldsymbol{m}\in \mathcal{D}^{'}\otimes \mathcal{L}$, means that all the elements of $\boldsymbol{m}$ (means, $m_{ij}$) are distributions of $\mathcal{D^{'}}$ type. Regarding the operations of Fourier transform and convolution between vector or matrix valued distributions, we recall that the Fourier transform of either forms is the Fourier transform of each of the elements and the convolution between two matrix valued distributions happens similar to the multiplication of two matrices while this time instead of simple multiplication, convolution does the job and this can be done similarly for the convolution between a matrix and a vector. 

%\subsection{Passivity and Herglotz}
\subsection{Immittance and scattering forms of passivity}
%{\color{blue} UNDER CONSTRUCTION...}
For a physical system with an input output relation in the convolution form, the requirement that the system be passive (output energy cannot exceed input energy) automatically implies that the system is causal as well \cite{srivastava2020causality}. The statement of passivity can be framed in two equivalent forms -- scattering and immittance. Consider, for example, an input-output relationship $\boldsymbol{x}=\boldsymbol{g}*\boldsymbol{f}$, whose passivity condition is given by the following scattering form:
\begin{equation}
\label{ePassive}
\int_{-\infty}^{t}\left(\boldsymbol{f}^\dagger\boldsymbol{f}-\boldsymbol{x}^\dagger\boldsymbol{x}\right)d{t'}\geq0,\;\forall\;t
\end{equation}
where $\dagger$ represents a conjugate transpose operation and $\boldsymbol{f}^\dagger\boldsymbol{f}$ is indicative of the $L_2$ energy in the input at time $t$. The above statement says that the total energy consumed in generating the output at any time $t$ can never exceed the total energy in the input to the system up to that time.

The passivity relations can be framed in another form, called the immittance form, which emerges naturally in certain problems. The introduction of new variables $\boldsymbol{v}(t)=\boldsymbol{f}(t)+\boldsymbol{x}(t)$ and $\boldsymbol{j}(t)=\boldsymbol{f}(t)-\boldsymbol{x}(t)$ allows us to write the passivity conditions as:
\begin{eqnarray}\label{ePassiveiv}
\Re\int_{-\infty}^{t}\boldsymbol{v}^\dagger \boldsymbol{j}d{t'}\geq0,\;\forall\;t
\end{eqnarray}

The interpretation of both forms of passivity is that in a passive system, the net absorbed energy of the system is non-negative. If the net absorbed energy is zero, then the system under consideration is conservative but still admissible as a passive system. Using the statement of passivity in the immittance form and assuming that $\boldsymbol{v}=\boldsymbol{z}*\boldsymbol{j}$ applies, we reiterate important results from Zemanian\cite{zemanian1963n} which are relevant here. For real transfer functions $\boldsymbol{z}(t)$, Zemanian showed that the following are true in the right half ($s>0$):
% used the statement of passivity in \ref{immittance2} and showed that the Laplace transform of the kernel for an immittance problem has non-negative definite Hermitian part. The recent property, along with being analytic in the right half plane and having real values for real amounts of argument in Laplace domain all together recalls a type of matrices called positive-real. The positive real matrices in Laplace domain ($\boldsymbol{W}(p)$) are those which for $Re[p]>0$ satisfy:
\begin{itemize}
    \item $\boldsymbol{\hat{z}}(p)$ is holomorphic
    \item $\boldsymbol{\hat{z}}(p^*)=\boldsymbol{\hat{z}}^*(p)$
    \item $\boldsymbol{\hat{z}}^\dagger(p) + \boldsymbol{\hat{z}}(p)$ is non-negative definite
\end{itemize}

For the scattering formalism, assuming that the input-output relationship is $\boldsymbol{x}=\boldsymbol{g}*\boldsymbol{f}$, the following results are true in the region $s>0$~\cite{beltrami1967linear}:
\begin{itemize}
    \item $\boldsymbol{\hat{g}}(p)$ is analytic/holomorphic.
    \item $\boldsymbol{\hat{g}}(p^{*})=\boldsymbol{\hat{g}}^{*}(p)$ 
    \item $\boldsymbol{I}-\boldsymbol{\hat{g}}^{\dagger}(p)\boldsymbol{\hat{g}}(p)$ is non-negative definite.
     
\end{itemize}
where $\boldsymbol{I}$ is the identity operator in the appropriate dimension. In the above, $\boldsymbol{\hat{z}}$ is called a positive real matrix whereas $\boldsymbol{\hat{g}}$ is called a bounded-real matrix~\cite{beltrami1967linear} and the two concepts are closely related to each other. In the preceding section, an integral representation was introduced for functions which have non-negative imaginary parts and are holomorphic in the upper half plane -- these are Herglotz functions. Functions possessing similar properties in the right half are called positive functions and when such functions represent real transfer functions in the time domain, then they are called positive real functions. Positive real functions are the scalar analogues of positive real matrices which were discussed in this section under the context of passivity in the immittance form. As mentioned earlier, the two concepts -- Herglotz functions/matrices and positive functions/matrices -- are connected to each other through a simple mapping. The former is defined in terms of $k=\omega+is$ whereas the latter is defined in terms of $p=s+i\omega$. Thus, it becomes evident that for an input-output relationship whose passivity statement may be written in the immittance form, the Laplace transform of its transfer is closely related to a Herglotz matrix. The connection is the following: for a time domain matrix $\boldsymbol{q}(t)$, we define its Laplace transform as either $\boldsymbol{Q}(p)=\left(q(t),e^{-pt}\right)$ or $\boldsymbol{Q}(k)=\left(q(t),e^{ikt}\right)$. If $\boldsymbol{Q}(p)$ is a positive matrix then $i\boldsymbol{Q}(k)$ is a Herglotz matrix.

\subsection{Herglotz matrix}
A positive real matrix with properties mentioned in the last sub-section has an integral representation\cite{zemanian1963n}:
\begin{equation}
\label{posrealmatrix}
    \boldsymbol{\hat{W}}(p)=-i\boldsymbol{C}+\boldsymbol{D}p+\int_{-\infty}^{+\infty}\frac{p}{p^{2}+r^{2}}(1+r^2)d\boldsymbol{\nu}(r)+\int_{-\infty}^{+\infty}\frac{1-p^2}{p^{2}+r^{2}}(r)d\boldsymbol{L}(r)
\end{equation}
In this representation, $\boldsymbol{C}$ is an $n\times n$ skew-symmetric matrix with pure imaginary elements, $\boldsymbol{D}$ is an $n\times n$ non-negative definite symmetric matrix with real elements, $\boldsymbol{\nu}$ is a symmetric matrix with real elements with bounded variations which are odd functions of $r$, and $\boldsymbol{L}$ is a skew-symmetric matrix with real elements with bounded variations which are even functions of $r$. The above representation for positive real matrices can be transformed into a representation for Herglotz matrices through a simple mapping. To be more specific, following the way used in Beltrami\cite{beltrami1966} to derive the integral representation for positive real matrices, one can apply the introduced mappings through the steps and find an integral representation for matrices which possess the properties of Herglotz like:

% Going through a similar process, one can show that a matrix which gets matched with all the properties of a positive real matrix, but this time it is holomorphic in the upper half plane has a similar integral representation which one can call it the tensorial counterpart of symmetric Herglotz representation. We recall that, the matrix $\boldsymbol{W}(k)$ in the following and the mentioned $\boldsymbol{W}(p)$ are both Laplace transforms of the same matrix in time domain which this time it is transformed by $F(k)=\left(f(t),e^{ikt}\right)$ (for each element). The representation would be like:
\begin{equation}
\label{tensorialHerglotz}
    \boldsymbol{\hat{W}}(k)=\boldsymbol{A}-i\boldsymbol{B}k+\int_{-\infty}^{+\infty}\frac{1+k^2}{r^{2}-k^2}rd\boldsymbol{\nu}(r)+i\int_{-\infty}^{+\infty}\frac{1+r^2}{k^{2}-r^{2}}k d\boldsymbol{L}(r)
\end{equation}
Here, $\boldsymbol{A}$ is a symmetric $n\times n$ matrix with real elements and $\boldsymbol{B}$ is a non-negative definite and skew-symmetric matrix with real members. $\boldsymbol{\nu}$ ($\boldsymbol{L}$) is a symmetric (skew-symmetric) matrix whose elements are real with bounded variations and even (odd) functions of $r$. \par
We can now take a similar set of steps as we took in section III and check whether the inverse transform of Eq. (\ref{tensorialHerglotz}) is a finite order derivative of a continuous matrix. Here, we do not reiterate the steps and just mention the inverse Fourier transform of the boundary value of Eq. (\ref{tensorialHerglotz}) in the $\mathcal{S}^{'}$ topology:

\begin{multline}
\label{tensorialHerglotzintime}
    \left(\boldsymbol{w}(t),\tilde{\phi}(t)\right)=\left(\boldsymbol{A}\delta(t),\tilde{\phi}(t)\right)-\left(i\boldsymbol{B}\delta(t)^{(1)},\tilde{\phi}(t)\right)-\left(\int\sin(rt)\theta(t)d\boldsymbol{\nu}(r),\tilde{\phi}^{(2)}(t)\right) \\+\left(\int\sin(rt)\theta(t) d\boldsymbol{\nu}(r),\tilde{\phi}(t)\right)+\left(\int\cos(rt)\theta(t) d\boldsymbol{L}(r),\tilde{\phi}(t)\right)\\-\left(\int\cos(rt)\theta(t)d\boldsymbol{L}(r),\tilde{\phi}^{(2)}(t)\right)+\left(\int\theta(t) d\boldsymbol{L}(r),\tilde{\phi}^{(2)}(t)\right)
\end{multline}
Thus we get:
\begin{equation}
\label{tensorialHerglotzintime2}
    \boldsymbol{w}(t)=\boldsymbol{A}\delta(t)-i\boldsymbol{B}\delta(t)^{(1)}+\frac{d^2}{dt^2}\left[\boldsymbol{P}_{2}(t)\theta(t)-\boldsymbol{P}_{1}(t)\theta(t)+\boldsymbol{P}_{1}(0)\theta(t)\right]+\boldsymbol{P}_{2}(t)\theta(t)+\boldsymbol{P}_{1}(t)\theta(t)
\end{equation}
Where, $\boldsymbol{P}_2(t)=\int_{-\infty}^{\infty}\sin(\xi t)d\boldsymbol{F}(\xi)$ and $\boldsymbol{P}_1(t)=\int_{-\infty}^{\infty}\cos(\xi t)d\boldsymbol{F}(\xi)$. It can be seen that all the non-zero terms in this equation are real and it is a second order derivative of two matrices with whose elements are continuous functions. Thus $l=2$ applies here as well which allows us to write the appropriate dispersion relation for $\boldsymbol{\hat{W}}(\omega)$ as:
\begin{eqnarray}
\label{dispersionW}
\boldsymbol{\hat{W}}(\omega)=-\frac{\omega^2}{\pi i}\left[\frac{\boldsymbol{\hat{W}}(\omega)}{\omega^2}*\mathcal{P}\left(\frac{1}{\omega}\right)\right].
\end{eqnarray}

% In other words, according to the mappings shown in the appendix A, the representation of a Herglotz function with the argument letter $k$, gets mapped to a positive function via the mapping relation of $k=ip$.

%From this point, we consider bold symbols as matrices/vectors and generalize the discussion to tensorial order as a scalar is a zero rank tensor.

\section{Relation between $\boldsymbol{C,D,L},B,\boldsymbol{\rho}$ and Herglotz functions - an appeal to passivity.}\label{sec:herglotzPassive}

The results in sections III and IV allow us to derive the correct order generalized Hilbert pairs for the material and metamaterial properties in acoustics and elastodynamics. 
% In other words, considering any of the material properties in either a function or tensorial form as transfer functions/matrices, one can design appropriate physical problems to first evaluate whether they are positive/positive real (or equivalently Herglotz/symmetric Herglotz) functions/matrices and then write the appropriate dispersion relations. It is also probable that instead of the transfer function itself an expression containing that transfer function be one of those types. 
Consider, for example, the case of elastodynamics whose equation of motion is given by:
% In this section, we review the results from the work done by Srivastava\cite{srivastava2015causality} to find a way to write an integral representation for transfer matrices in an elastodynamic problem. We begin this review by mentioning equations of motion for a general elastodynamics problem like:
\begin{equation}
    \label{EOM1}
    \sigma_{ij,j}+f_i=\dot{p}_{i},
\end{equation} 
where $\boldsymbol{\sigma}$, $\boldsymbol{f}$ and $\boldsymbol{p}$ are space and time dependant representations of stress, external force and momentum respectively. This equation of motion is augmented with constitutive relations which, in the linear regime, are given by~\cite{srivastava2015causality}:
% To write the input-output relation for this problem by considering linearity, single-valuedness and time-invariance, we consider stress and velocity as input fields and strain and momentum as the response to them and recall the constitutive relations as:
\begin{equation}
\begin{aligned}
    \label{constitutive1}
    \boldsymbol{\epsilon}(\boldsymbol{x},t)=\boldsymbol{D}*\boldsymbol{\sigma}=\int\boldsymbol{D}(\boldsymbol{x},t-\tau):\boldsymbol{\sigma}(\boldsymbol{x},\tau)d\tau,\\
        \boldsymbol{p}(\boldsymbol{x},t)=\boldsymbol{\rho}*\dot{\boldsymbol{u}}=\int\boldsymbol{\rho}(\boldsymbol{x},t-\tau).\dot{\boldsymbol{u}}(\boldsymbol{x},\tau)d\tau,
\end{aligned}
\end{equation}
where $\boldsymbol{\epsilon},\dot{\boldsymbol{u}}$ are strain and velocity respectively. In the above, $\boldsymbol{D}$ is the time domain compliance tensor and $\boldsymbol{\rho}$ is the time domain density tensor. The field variables and the constitutive tensors are all assumed to belong to appropriate distribution spaces (see~\cite{srivastava2015causality} for details). 
% For convenience in next steps, we assume that the matrix distributions $\boldsymbol{D}$ and $\boldsymbol{\rho}$ are members of $\mathcal{S}^{'}_{n\times n\times n\times n}$ and $\mathcal{S}^{'}_{n\times n}$ respectively and both have real elements in the time domain. It is also enough to consider $\boldsymbol{\sigma}$ and $\dot{\boldsymbol{u}}$ as the members of $\mathcal{D}_{n\times n}$ and $\mathcal{D}_{n\times 1}$ respectively. 
The total absorbed energy at time $t$ in a system characterized by Eqs. (\ref{EOM1},\ref{constitutive1}) and occupying a region $\Omega$ can be derived as:
\begin{equation}
    \label{passivity1}
    E(t)=\int_{-\infty}^{t}\frac{\partial E(\tau)}{\partial \tau}d\tau=\Re\frac{1}{2}\int_{-\infty}^{t}d\tau\int_{\Omega}d\boldsymbol{x}\frac{\partial}{\partial \tau}\left[\boldsymbol{\sigma}(\boldsymbol{x},\tau):\boldsymbol{\epsilon}^{*}(\boldsymbol{x},\tau)+\boldsymbol{p}(\boldsymbol{x},\tau).\dot{\boldsymbol{u}}^{*}(\boldsymbol{x},\tau)\right].
\end{equation}
Passivity implies that $E(t)\geq 0\;\forall t$. Since the absorbed energy $E(t)$ is related to the power $P(t)$ through $P(t)=dE/dt$ and since the power $P(t)$ may be expressed in terms of the work done by the body forces $\boldsymbol{f}$ and surface tractions $\boldsymbol{t}$, one can eventually arrive at the following relation implied by passivity~\cite{srivastava2015causality}:
% Recalling that the total absorbed energy by the system ($E(t)$) should be non-negative at any an moment, We can re-write the recent expression by noting that the time derivative of $E(t)$ is equal to the power input from the tractions on the boundary and body forces on $\Omega$ itself like:
\begin{equation}
    \label{passivity2}
\Re\int_{-\infty}^{t}ds\left[\sigma_{ij}(\tau)\dot{\epsilon}^{*}_{ij}+\dot{p}_{i}(\tau)\dot{u}^{*}_i\right]\geq 0
\end{equation}
One can view the above relation as an example of the immittance form of passivity. By employing the constitutive relations (\ref{constitutive1}) and keeping in mind that the real part of the equation above is being taken, one can show that the passivity equation implies the following conditions:
% To stay committed to the notation about compliance tensor, we showed its time domain representation in upper case letter while due to the introduced convention in this paper the time domain tensors should be represented by small letters. Here, we define $\hat{(.)}$ as the Laplace transform, i. e. ($\hat{f}(p)=(f,e^{-pt})$) or $\hat{f}(k)=\left(f(t),e^{ikt}\right)$. Considering that the input elements are independent of each other, we define $\boldsymbol{\sigma}(\tau)=\boldsymbol{\sigma}\phi^{*}(\tau)$ and $\dot{\boldsymbol{u}}(\tau)=\dot{\boldsymbol{u}}\gamma(\tau)$ where $\boldsymbol{\sigma}$ and $\dot{\boldsymbol{u}}$ are constant tensors and $\phi$ and $\gamma$ are members of spaces of test functions, $\mathcal{D}$. Now, depending on the choice of testing functions we can investigate the effect of passivity on the Laplace transform of tensors. For instance, choosing $\phi$ and $\gamma$ both as $e^{p\tau}$ (corresponding to the Laplace transform formula, $\hat{f}(p)=\left(f(t),e^{-pt}\right)$) after some manipulations and rearrangements gives:
\begin{equation}
\begin{aligned}
    \label{passivity4}
\hat{\dot{\boldsymbol{D}}}^{h}(p),\hat{\dot{\boldsymbol{\rho}}}^{h}(p)\geq 0
% \boldsymbol{\phi}:\hat{\dot{\boldsymbol{D}}}^{h}:\boldsymbol{\phi}^{*}\geq 0\\
% \boldsymbol{q}:\hat{\dot{\boldsymbol{\rho}}}^{h}:\boldsymbol{q}^{*}\geq 0
\end{aligned}
\end{equation}
In the above, the hat represents the Laplace transform, the superscript $h$ ($nh$) represents the hermitian (non-hermitian) part of the tensor, and the inequality is understood in terms of positive semi-definiteness. Thus, passivity implies that $\hat{\dot{\boldsymbol{D}}},\hat{\dot{\boldsymbol{\rho}}}$ are positive tensors if the Laplace transform is defined with respect to the parameter $p$. Consequently, $i\hat{\dot{\boldsymbol{D}}}(k),i\hat{\dot{\boldsymbol{\rho}}}(k)$ are Herglotz tensors. This has further consequences due to the general Laplace transform relation $\mathcal{L}\left[f^{(n)}(t)\right]=(-ik)^{(n)}\hat{f}(k)$. For example, from $\hat{\dot{\boldsymbol{D}}}=-ik\hat{{\boldsymbol{D}}}$, we get $k\hat{{\boldsymbol{D}}}=i\hat{\dot{\boldsymbol{D}}}$. Since $i\hat{\dot{\boldsymbol{D}}}(k)$ has a positive semi-definite non hermitian part on account of it being Herglotz, it implies that $k\hat{{\boldsymbol{D}}}$ also has a positive semi definite non hermitian part. As a corollary, for real frequency $\omega$ and a scalar compliance $D$, this result means that the imaginary part of $\omega\hat{D}$ must be a non negative quantity due to passivity. Similar results hold for $\hat{\boldsymbol{\rho}}$ as well. Since $i\hat{\dot{\boldsymbol{D}}}(k),i\hat{\dot{\boldsymbol{\rho}}}(k)$ are Herglotz tensors, it automatically means that the lowest order dispersion relations that one can write on $i\hat{\dot{\boldsymbol{D}}}(k),i\hat{\dot{\boldsymbol{\rho}}}(k)$ is order 2 ($l=2$). Therefore, we have the following dispersion relations:

\begin{equation}
\begin{aligned}
    \label{dispersionDrho}
    \omega\hat{\boldsymbol{D}}(\omega)=-\frac{\omega^2}{\pi i}\mathcal{P}\int_{-\infty}^{+\infty}\frac{\omega'\hat{\boldsymbol{D}}(\omega')}{\omega'^{2}(\omega-\omega')}d\omega'\\
    \omega\hat{\boldsymbol{\rho}}(\omega)=-\frac{\omega^2}{\pi i}\mathcal{P}\int_{-\infty}^{+\infty}\frac{\omega'\hat{\boldsymbol{\rho}}(\omega')}{\omega'^{2}(\omega-\omega')}d\omega'
\end{aligned}
\end{equation}
After some algebraic manipulations and using the symmetry relations $\hat{\boldsymbol{D}}(-\omega)=\hat{\boldsymbol{D}}^{*}(\omega)$ and $\hat{\boldsymbol{\rho}}(-\omega)=\hat{\boldsymbol{\rho}}^{*}(\omega)$, the dispersion relations in Eq. (\ref{dispersionDrho}) can be re-written only for positive frequencies:
\begin{equation}
\begin{aligned}
    \label{dispersionDrho2}
    \Re\hat{\boldsymbol{D}}(\omega)=-\frac{\omega}{\pi}\mathcal{P}\int_0^\infty\frac{\omega\Im\hat{\boldsymbol{D}}(\omega')}{\omega'(\omega^{2}-\omega'^{2})}d\omega';\quad
    \Im\hat{\boldsymbol{D}}(\omega)=\frac{\omega}{\pi}\mathcal{P}\int_0^\infty\frac{\omega'\Re\hat{\boldsymbol{D}}(\omega')}{\omega'(\omega^{2}-\omega'^{2})}d\omega'
\end{aligned}
\end{equation}
for the real and imaginary parts of $\hat{\boldsymbol{D}}$, and:

\begin{equation}
\begin{aligned}
    \label{dispersionDrho3}
    \Re\hat{\boldsymbol{\rho}}(\omega)=-\frac{\omega}{\pi}\mathcal{P}\int_0^\infty\frac{\omega\Im\hat{\boldsymbol{\rho}}(\omega')}{\omega'(\omega^{2}-\omega'^{2})}d\omega';\quad 
    \Im\hat{\boldsymbol{\rho}}(\omega)=\frac{\omega}{\pi}\mathcal{P}\int_0^\infty\frac{\omega'\Re\hat{\boldsymbol{\rho}}(\omega')}{\omega'(\omega^{2}-\omega'^{2})}d\omega'
\end{aligned}
\end{equation}
for the real and imaginary parts of $\hat{\boldsymbol{\rho}}$. In addition to the above, there are some further interesting conclusions on the analytic structure of the inverse of the compliance tensor -- the stiffness tensor $\hat{\boldsymbol{C}}=\hat{\boldsymbol{D}}^{-1}$ -- due to the fact that the inverse of a positive real matrix is also positive real\cite{mcmillan1952introduction}. 
%{\color{red}I think the conclusion below does not necessarily follow - maybe a different argument needs to be made. $1/kD$ is Herglotz, therefore, $C/k$ is Herglotz not $kC$. This would seem to mean that $C$ will have $l=3$! What happens when you try to express Eq. 30 in terms of C? Do you still get $l=3$?}

Since $p\hat{\boldsymbol{D}}(p)$ is a positive real tensor, it follows that $1/p\hat{\boldsymbol{D}}(p)$ or $\hat{\boldsymbol{C}}(p)/p$ is also a positive real tensor. As a corollary, it also follows that $-\hat{\boldsymbol{C}}(k)/k$ is Herglotz. At this point, the positive realness of $\hat{\boldsymbol{C}}(p)/p$ immediately means that the lowest order dispersion relation for $\hat{\boldsymbol{C}}$ is $3$ ($l=3$). This result is also provable by employing the passivity statement in the immittance form, this time by assuming the strain tensor, $\boldsymbol{\epsilon}$, as the input (see Appendix II).
%{\color{red}Is this statement correct?}{\color{blue}You are right, my statement is not right. This time by doing the Laplace with respect to $k$ in the passivity statement,  we get $\Re\left[\frac{i\hat{\boldsymbol{C}}(k)}{k}\right]$ which means that $i\left(\frac{\hat{\boldsymbol{C}}(k)}{k}\right)^{nh}\geq 0$ which this means that the imaginary part of $\frac{\hat{\boldsymbol{C}}(k)}{k}$ is negative so it is not Herglotz. Does that make sense? I think everything is right but I was expecting to get that C/k is also Herglotz. At this point we can conclude that as $\frac{1}{i}\left(\frac{\hat{\boldsymbol{C}}(k)}{k}\right)^{nh}< 0$ so one can say that $-\frac{\hat{\boldsymbol{C}}(k)}{k}$is Herglotz which I do not know if there is any specific interpretation for this result. Please see the appendix II. }
%{\color{red}Add a half page appendix; no need to present Approach 1 and 2. Just do the one where C/p comes out directly as positive.}). 
% By doing so, one can reach to the conclusion that the real part of $\frac{1}{p}\hat{\boldsymbol{C}}(p)$ is a non-negative definite matrix which in conjunction with the analyticity in the right half plane and having the same symmetry as $\hat{\boldsymbol{D}}(p)$ means that $\frac{1}{p}\hat{\boldsymbol{C}}(p)$ is a positive real matrix. 
Thus, the lowest order dispersion relations applicable to $\hat{\boldsymbol{C}}(\omega)$, based purely upon passivity, are:

\begin{equation}
\begin{aligned}
    \label{dispersionC}
    \Re\hat{\boldsymbol{C}}(\omega)=-\frac{\omega^{3}}{\pi}\mathcal{P}\int_{0}^{\infty}\frac{\omega\Im\hat{\boldsymbol{C}}(\omega')}{\omega'^{3}(\omega^{2}-\omega'^{2})}d\omega';\quad 
    \Im\hat{\boldsymbol{C}}(\omega)=\frac{\omega^{3}}{\pi}\mathcal{P}\int_{0}^{\infty}\frac{\omega'\Re\hat{\boldsymbol{C}}(\omega')}{\omega'^{3}(\omega^{2}-\omega'^{2})}d\omega'
\end{aligned}
\end{equation}
Some important points need to be made here. Our analysis shows that passivity in elastodynamics implies that $\omega\hat{\boldsymbol{D}}(\omega)$ is Herglotz and in this conclusion, the compliance tensor is analogous to the electrical permittivity tensor ($\boldsymbol{\epsilon}$) of electromagnetism which also has the same Herglotz behavior\cite{sumrules}. Bernland et al.\cite{sumrules} have shown that $\omega\boldsymbol{\epsilon}(\omega)$ is Herglotz but then they go on to derive a zero order dispersion relation on $\boldsymbol{\epsilon}(\omega)-\boldsymbol{\epsilon}(\infty)$. However, it must be noted that in doing so, they assume further restrictions on the behavior of $\boldsymbol{\epsilon}(\omega)$ than asserted merely by passivity. These include not only continuity and boundedness but also that $\boldsymbol{\epsilon}(\omega)-\boldsymbol{\epsilon}(\infty)$ goes down as $\mathcal{O}(1/\omega)$ as $|\omega|\rightarrow\infty$. Similarly, Muhelstein et al.~\cite{muhlestein2016reciprocity} assert a zero order dispersion relation on the Willis tensor. It is likely that similar constraints are implied in their results as well but they are not explicitly clarified. However, it should be noted that these constraints, while reasonable, follow from neither passivity nor causality. Passivity implies causality and, by itself, only implies that order 1 dispersion relations apply to the compliance tensor of elastodynamics and the electrical permittivity tensor of electromagnetism.

% Based on our conclusion that $k\hat{\boldsymbol{D}}$ is a Herglotz matrix which, one can conclude that for the case in which $D$ is a scalar (1D elastic problem), it does not have zeros in the upper half plane which immediately means that the inverse of $D$ (Young's modulus, $E$) has no pole in the same half-plane. Thus, passivity ensures that both $D$ and $E$ are causal transforms.

Similar to the treatments applied to a general elastodynamic case, a slightly simpler example can be considered which allows one to evaluate the effect of passivity on the bulk modulus $B$. 
% Due to the point that Bulk modulus points the relation between volumetric stress and strain, in this case the statement of passivity can be written for a non-viscose fluid element under normal pressure to the surfaces of the elements. To repeat a set of similar manipulations like the general elastodynamic case, 
We first express Eq. (\ref{passivity2}) in an alternate form involving the deviatoric and volumetric parts of stress and strain\cite{achenbach1984wave}:
\begin{equation}
\label{Bulk1}
    \sigma_{ij}\dot{\epsilon}_{ij}^{*}=\left(s_{ij}-M\delta_{ij}\right)\left(\dot{e}_{ij}+\frac{\dot{d}}{3}\delta_{ij}\right)^{*}
\end{equation}
Here, $s_{ij}$ is the deviatoric part of the stress tensor and $M=-\frac{\sigma_{kk}}{3}$ is the volumetric part. Similarly, $\dot{e}_{ij}$ is the first derivative of the deviatoric part of the strain tensor and $\dot{d}=\dot{\epsilon}_{kk}$ is the time derivative of the volumetric part. At this point, the passivity statement (\ref{passivity2}) can be employed only keeping the volumetric effects. By employing the constitutive relation between $d$ and $M$, $M(t,\boldsymbol{x})=-B(t,\boldsymbol{x})\ast d(t,\boldsymbol{x})$, and considering velocity and the volumetric strain separately as the inputs to the system, we arrive at the result that the Bulk modulus behaves in a similar fashion as the stiffness tensor and the inverse of the Bulk modulus has the same behavior as the compliance tensor. More precisely, $l=3$ for $\hat{B}$ and $l=1$ for $1/\hat{B}$.

The above analysis can be extended to materials characterized by passive linear Willis tensors -- a class of constitutive relations which encompasses both the elastodynamic and acoustic cases \cite{willis1997dynamics,GalforWillis,muhlestein2016reciprocity,nemat2011homogenization,alizadeh2021overall,aghighi2019low,amirkhizi2018overall,Norris1,nassar2020nonreciprocity,nassar2017modulated,chen2020active,muhlestein2017experimental,muhlestein2017analysis,PhysRevB.96.104303}. The details of the following analysis are given in Srivastava~\cite{srivastava2015causality} and here we outline only the most pertinent points. Linear Willis materials are those for which the constitutive relations exhibit a coupled form. Defining the stress and the velocity as the inputs (represented in the vector $\boldsymbol{w}(t)$) and the strain and momentum as the outputs (represented in the vector $\boldsymbol{v}(t)$) of the system, and noting that the elements of the input vectors are assumed to be in $\mathcal{D}$, one can write a single convolution relation for a linear, real, time-invariant and causal Willis material:
\begin{equation}
    \label{Willis1}
    \boldsymbol{v}(t)=\int_{-\infty}^{\infty}\boldsymbol{L}(t-s)\boldsymbol{w}(s)ds
\end{equation}
In Eq. (\ref{Willis1}), $\boldsymbol{v}=\begin{pmatrix}
\boldsymbol{\epsilon} \\
\boldsymbol{p}
\end{pmatrix}$, $\boldsymbol{w}=\begin{pmatrix}
\boldsymbol{\sigma} \\
\dot{\boldsymbol{u}}
\end{pmatrix}$. Moreover, the kernel of the integral in Eq. (\ref{Willis1}), is a $n\times n$ matrix whose elements are distributions of slow growth. At this point, the requirement enforced by the passivity can be represented like:
\begin{equation}
    \label{Willis2}
    E(t)=\Re\int_{-\infty}^{t}\boldsymbol{w}^{\dagger}(s)\dot{\boldsymbol{v}}(s) ds\geq 0
\end{equation}
By applying the convolution relation (\ref{Willis1}), one can write:
\begin{equation}
    \label{Willis3}
    E(t)=\Re\int_{-\infty}^{t}ds\boldsymbol{w}^{\dagger}(s)\int_{-\infty}^{\infty}\dot{\boldsymbol{L}}(\nu)\boldsymbol{w}(s-\nu)d\nu\geq 0
\end{equation}
where,
\begin{equation*}
    \label{L}
\boldsymbol{L}=
\begin{pmatrix}
\boldsymbol{D} & \boldsymbol{S}_{1} \\
\boldsymbol{S}_{2} & \boldsymbol{\rho}
\end{pmatrix}
\end{equation*}
is the coupled Willis constitutive tensor. In the above, $\boldsymbol{D}$ and $\boldsymbol{\rho}$ are compliance and density tensors and $\boldsymbol{S}_1$ and $\boldsymbol{S}_2$ are coupling tensors. Employing a similar method which was used for materials which are not of Willis type to take the integrands to the Laplace domain, one can conclude that passivity implies:
\begin{equation}
    \label{Willis4}
    \boldsymbol{y}^{\dagger}\hat{\dot{\boldsymbol{L}}}^{h}(p)\boldsymbol{y}\geq 0
\end{equation}
The rest of the analysis follows from the previous concerns. Specifically, we arrive at the conclusion that $p\hat{\boldsymbol{L}}(p)$ is a positive real tensor and that $k\hat{\boldsymbol{L}}(k)$ is Herglotz. Thus, it follows that $l=1$ applies to $\hat{\boldsymbol{L}}$ and $l=3$ applies to $\hat{\boldsymbol{L}}^{-1}$.

\section{Dispersion relations for $\kappa,n'$}\label{sec:herglotznZkappa}

\begin{figure}[!htb]
\includegraphics[scale=0.3]{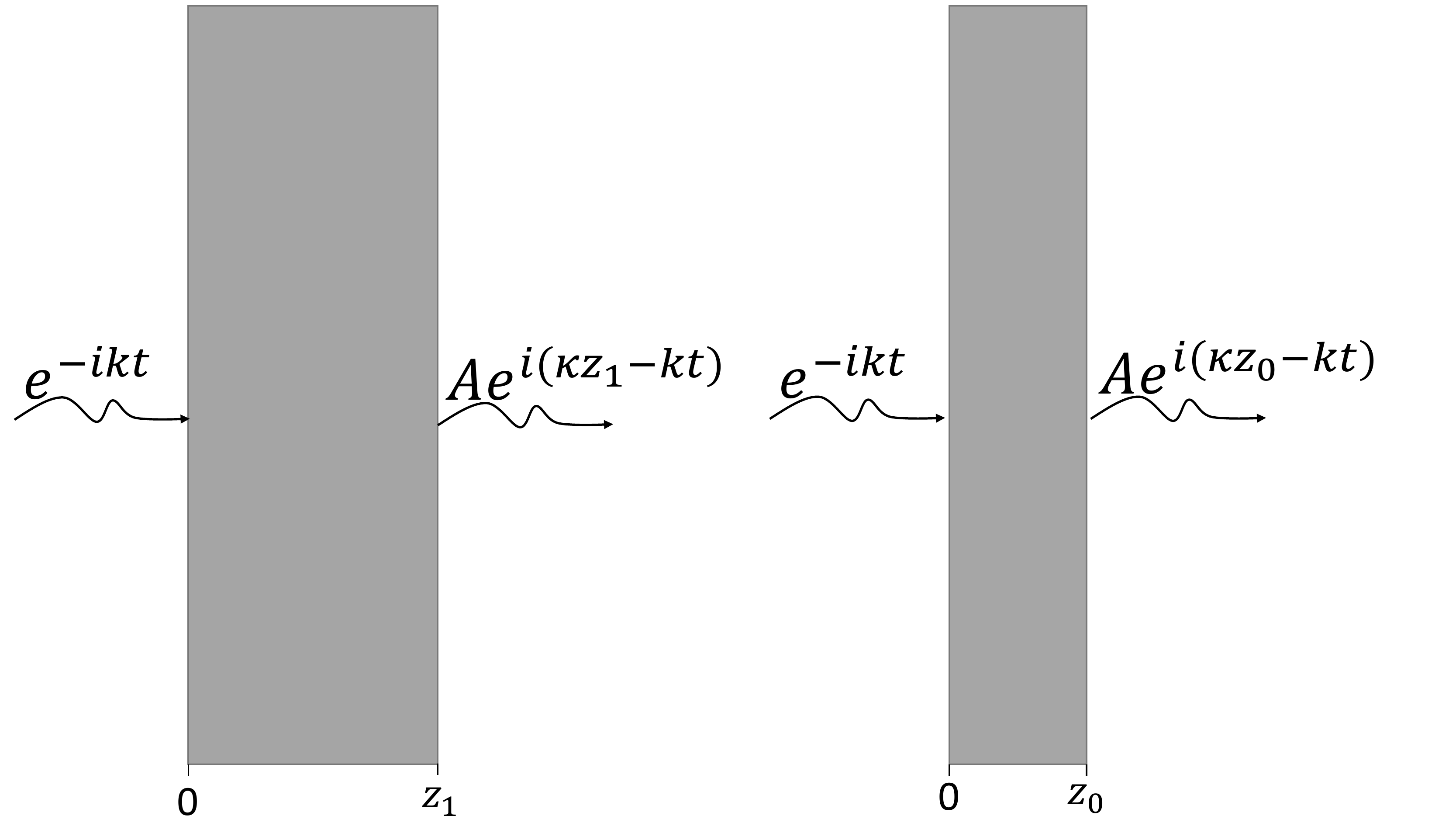}
\centering
\caption{ Schematic of the propagation of a plane wave through a thin slab.}
\centering
\label{Fig2}
\end{figure}

Appropriate dispersion relations may also be derived for the wavenumber $\kappa$ and the refractive index $n'$ which emerge in wave propagation problems in acoustics and elastodynamics, however, the treatment does not parallel the one which we employed for constitutive tensors. 
% Following the point that all the functions/matrices discussed in previous sections were the kernels of operators connecting two physical quantities, in this section we discuss two more material or metamaterial properties which are not necessarily pointing the connection between a pair of physical input-output quantities.
% Causality and dispersion relation discussions on such functions are a little less straightforward as both of the passivity statements used for previous transfer functions/matrices are explained based upon the input-output relations. 
Analyticity and non-negative imaginary part in the upper half-plane for $\kappa(k)$ were shown indirectly by Weaver and Pao~\cite{weaver1981dispersion} by relying on the analyticity of the Green's function in a simple wave propagation problem (Fig. \ref{Fig2}). In this problem, an input plane wave given by $\hat{f}(k)=e^{-ikt}$ is incident on a slab at $z=0$. As it travels a distance $z$ in the slab, it is transformed into a form $\hat{u}(k)=A(k)e^{i(\kappa z-kt)}$ where $A(k)$ encapsulates both the phase and amplitude modifications of the wave as it travels in the slab. The relation between $\hat{u}(k)$ and $\hat{f}(k)$ is, therefore, $\hat{u}(k)=A(k)e^{i\kappa z}\hat{f}(k)$ where $A(k)e^{i\kappa z}\equiv \hat{g}(k,z)$ is the Green's function of the problem. Weaver and Pao used two slabs made of the same material (Fig. \ref{Fig2}) but of thicknesses $z_0,z_1=z_0+d$ to arrive at the relation $\kappa=\frac{-i}{d}\ln{\frac{\hat{g}(k,z_1)}{\hat{g}(k,z_0)}}$. They showed that, given the analyticity of the Green's function, the analyticity of the wavenumber follows. In addition, passivity requires that the energy contained in the wave, as it travels through the slab, must be monotonically non-decreasing. This results in a passivity statement in the scattering form:

\begin{equation}
    \label{passivitywave}
    \int_{-\infty}^{t}\left(f^{*}f- u^{*}u\right)d{t'}\geq0,\;\forall\;t
\end{equation}
From this passivity statement, Weaver and Pao concluded that the wavenumber has a non-negative imaginary part in the upper half. Analyticity combined with the non-negative imaginary part in the upper half ensure that $\kappa$ is a Herglotz function but not necessarily a symmetric one, which allowed Weaver and Pao to derive a dispersion relation for it.

%One can also employ a result described in Bernland et al.\cite{gustafsson2010sum} who discussed that for passive systems in a scattering formalism, one can also define a Herglotz function. They showed that the Green's function, $G(k)$, in a simple wave propagation problem through a slab\cite{nussenzveig1972causality} can be expressed in a form $G(k)=b(k)e^{i\kappa(k)}$ {\color{red}Is there no $z$ here in the exponential? This is different from earlier.} where $b(k)$ is a Blaschke product and $\kappa(k)$ is a Herglotz function.

 Knowing that $\kappa$ is a Herglotz function and it is not necessarily symmetric immediately implies that the lowest dispersion relation that one can write for $\kappa$ is 2 ($l=2$):
\begin{equation}
    \label{dispersionkappa}
    \kappa(\omega)=-\frac{\omega^2}{\pi i}\mathcal{P}\int_{-\infty}^{+\infty}\frac{\kappa(\omega')}{\omega'^{2}(\omega-\omega')}d\omega'
\end{equation}

As a final point, we can consider the dispersion relations on the refractive index $n'$. For acoustic waves, for instance, the phase velocity $c_p$ is related to the Bulk modulus and density through the relation $c_p=\sqrt{\hat{B}/\hat{\rho}}$ and the wavenumber is related to the phase velocity as $\kappa=k/c_p$. Referring back to the analysis associated with Fig. (\ref{Fig2}), we now have the Green's function relationship 
$G(z,k)=A(k)e^{ikn'(k) z}$. Since the Green's function satisfies the passivity relationship in the scattering form, it follows that $kn'(k)$ is a Herglotz function\cite{weaver1981dispersion}. Therefore, we can conclude that the lowest dispersion relation order for $n'$ coming from passivity is $1$, ($l=1$):

%{\color{red}Provide the dispersion relation.}

\begin{equation}
    \label{dispersionn'}
    n'(\omega)=-\frac{\omega}{\pi i}\mathcal{P}\int_{-\infty}^{+\infty}\frac{n'(\omega')}{\omega'(\omega-\omega')}d\omega'
\end{equation}

\section{Conclusions}\label{sec:conclusions}

In this paper, we have derived dispersion relations for the material tensors and transfer functions of acoustic and elastodynamic materials and metamaterials. Generally this requires one to know the behavior of these properties in the limit $|\omega|\rightarrow \infty$, however, this information is not easy to ascertain. In fact, it may not even make sense to talk about this asymptotic limit as the continuum approximation breaks down in the high frequency regime. To sidestep this issue, we resort to the principle of passivity and its connection to Herglotz functions and positive functions. Our analysis in this paper concerns only passive systems -- systems which encompass no sources of energy. Since Herglotz functions have well known dispersion relations, our overarching aim in this paper is to relate the various transfer functions and material properties of passive acoustic and elastodynamic systems to Herglotz functions. However, for completeness sake, we also include some relevant and classical results.
% Following the discussions on the analyticity and dispersion relations of material properties, whether they act as a transfer function between two physical quantities or not, we set a theorem from the theory of distributions as the cornerstone of this paper and developed a set of dispersion relations with the minimum possible order for transfer functions in acoustics and elastodynamics. Earlier, the efforts on developing dispersion relations, which can be referred as bounds on material properties, were mostly dependant on some assumptions on the growth rate of the transfer functions in high frequency domain. 
We first clarify that the inverse Fourier transform of any Herglotz function is a second order derivative of a continuous function. This immediately establishes the classical result that dispersion relations of order 2 apply to Herglotz functions. Then we describe the immittance and scattering forms of passivity, especially clarifying the classical result which connects the transfer functions appearing in immittance forms of passivity to positive functions. We subsequently clarify the connection between positive functions and Herglotz functions. Thus, we clarify the connections between positive functions, Herglotz functions, passivity, and dispersion relations. These developments then allow us to derive the appropriate dispersion relations on wavenumber ($\kappa$), refractive index ($n$), density ($\boldsymbol{\rho}$) and its inverse, stiffness ($\boldsymbol{C}$) and compliance ($\boldsymbol{D}$) tensors, the Bulk modulus ($B$), and finally the broader generalization of these properties which is now known as the Willis tensor ($\boldsymbol{L}$). Our analysis shows that based upon passivity alone, dispersion relations of \emph{minimum} order 1 apply to the Fourier transforms of $\boldsymbol{D},\boldsymbol{\rho}, n'$, and the inverse of $B$, order 3 apply to $\boldsymbol{C},B$, and the inverse of $\boldsymbol{\rho}$, and order 2 applies to $\kappa$.

% Section IV presents a brief but substantial review on the forms of passivity statement, called scattering and immittance, and showed that the transfer matrix in Laplace domain is a positive real tensor if a physical system is passive in its immittance form and the transfer tensor in time domain has real valued elements. Section II, II and IV also make clear that positive realness and being Herglotz are necessarily pointing a single concept and they can be related to each other in some way. Applying the passivity statement for transfer functions which are relating two physical quantities like $\hat{\boldsymbol{C}}$, $\hat{\boldsymbol{D}}$, $\hat{\boldsymbol{\rho}}$ and $\hat{B}$ or going through an indirect way for functions which are not necessarily transferring a physical input to its corresponding out put like $n'$ and $\kappa$, assisted us to develop dispersion relations for all of the mentioned functions/matrices in sections V and VII. At the ened we recall that, the results of this paper clarify a set of more exact relations between real and imaginary parts of material properties in acoustics and elastodynamics by discussing them in a weak topology which is a major benefit of distribution theory.
\section{Appendix}\label{sec:appendix}
\subsection{Appendix I}
In this appendix, we present an introduction about an integral representation called "Herglotz" for two related functions -- those which are holomorphic in the upper half of the complex domain and have positive imaginary parts and those which are holomorphic in the right half plane and have positive real parts. Cauchy's integral formula is an integral representation which describes the value of a holomorphic function at a certain point in terms of an integral over the boundary of a closed contour containing that point. Assuming a function, $f(z)$, analytic in the unit circle in the complex plane, Cauchy's integral gives:
\begin{equation}
    \label{Cauchy1}
    f(z)=\frac{1}{2\pi i}\int_{|\xi|=1}\frac{f(\xi)}{\xi-z}d\xi
\end{equation}
If $f$ has a positive real part inside the unit circle, then it is possible to represent the integral formula in the following representation~\cite{Herglotzthesis}:

\begin{equation}
    \label{rep2}
    f(z)=i\Im\left(g\left(0\right)\right)+\frac{1}{2\pi }\int_{0}^{2\pi}\frac{e^{is}+z}{e^{is}-z}d\nu\left(s\right)
\end{equation}
where where  $\nu(s)$ on $U$ is called a Borel measure and $U=\left[0,2\pi \right)$. The measure is real and uniformly bounded. Eq. (\ref{rep2}) can be mapped to any of the half planes under discussion in this paper using conformal mappings. For instance, to obtain the so called Herglotz representation\cite{sumrules,cassier2017bounds}, we first map the function to the upper half plane and then define another transformation $h=if$. Since $f$ has a positive real part inside the unit circle, $h$ will have a positive imaginary part in the upper half. The composite mapping function which accomplishes this is:
\begin{equation}
    \label{map1}
    k\equiv i\frac{1+z}{1-z}
\end{equation}
where $k$ is a complex number, and the function, after the transformations is:
\begin{equation}
    \label{Herg1}
    h(k)=-\Im\left[f(0)\right]+\frac{1}{2\pi}\nu\left({0}\right)k+\frac{1}{\pi}\int_{-\infty}^{+\infty}\left(\frac{1}{r-k}-\frac{r}{1+r^2}\right)d\mu(r)
\end{equation}
where, $d\nu=\frac{2}{1+r^2}d\mu(r)$, $e^{is}=\frac{r-i}{r+i}$, and $-\infty<r<\infty$ is real. Equivalently, Eq. (\ref{Herg1}) can be re-written as:
\begin{equation}
    \label{Herg2}
    h(k)=a+bk+\frac{1}{\pi}\int_{-\infty}^{+\infty}\frac{1+rk}{r-k}d\nu(r)
\end{equation}
which is the usual Herglotz representation where $a$ is real, $b$ is non-negative and $\nu$ is a positive bounded measure. To get to the positive representation ($p-$representation), one can simply map the unit circle to the right half with:
\begin{equation}
    \label{map2}
    p\equiv \frac{1+z}{1-z}
\end{equation}
which yields~\cite{cauer1932poisson}:
\begin{equation}
    \label{righthalf}
    h(p)=c+dk+\frac{1}{\pi}\int_{-\infty}^{+\infty}\frac{irp-1}{ir-p}d\nu(r)
\end{equation}
where $d$ is non-negative and $c$ is a pure imaginary constant. Similar to the symmetric Herglotz representation, by considering the symmetry relation $h(\bar{p})=\bar{h}(p)$ (if $h(p)$ is real in its time domain), one can write:
\begin{equation}
    \label{posreal}
    h(p)=dp+\frac{1}{\pi}\int_{-\infty}^{+\infty}\frac{p\left(1+r^{2}\right)}{p^{2}+r^{2}}d\nu(r)
\end{equation}
which is the so called positive-real function representation, equivalent to the symmetric Herglotz representation.

\subsection{Appendix II}
In this appendix, following the discussion on the dispersion relation for the stiffness matrix, we review the application of the passivity statement in its immittance form when the strain tensor is the input instead of the stress tensor. To avoid mentioning unnecessary equations here, we immediately start from Eq. (\ref{passivity2}). Using the constitutive relation in the time domain, the first term of Eq. (\ref{passivity2}), can be re-written:
\begin{equation}
    \label{passivityC}
    \Re\int_{-\infty}^{t}ds\left[\dot{\epsilon}^{*}_{ij}(s)\int_{-\infty}^{+\infty}C_{ijkl}\epsilon_{kl}(s-v)dv\right]\geq 0
\end{equation}
Since we are not considering a restrictive problem in which the inputs are members of $L^{2}$, instead of using the Plancherel's theorem to transform the stiffness matrix to the Laplace domain, we go through the process which Zemanian\cite{zemanian1963n} and Srivastava\cite{srivastava2015causality} employed in their derivations. Recalling that the stiffness matrix here is considered to be a tempered distribution, to make sure that the Laplace transform is definable, the input $\boldsymbol{\epsilon}(s)$ is expressed as $\epsilon_{ij}(s)=\epsilon_{ij}\phi(s)$, where $\epsilon_{ij}$ is a constant matrix and $\phi(s) \in \mathcal{S}$. Since the members of $\mathcal{S}$ are infinitely differentiable and continuous, one can also define, $\dot{\epsilon}_{ij}(s)=\epsilon_{ij}\dot{\phi}(s)$. Substituting the introduced expressions for $\epsilon_{ij}$ and $\dot{\epsilon}_{ij}$ into Eq. (\ref{passivityC}), and by assuming $\phi(s)=e^{ps}$, we arrive at:
%{\color{red}Professor, I wrote this section using your paper and Zemanian which to form the Laplace transform of $C_{ijkl}$ we define $\phi$ like $e^{ps}$ and claim that it is a member of $\mathcal{S}$ which are fast decreasing functions but I cannot imagine $e^{ps}$ or $e^{-iks}$ as fast decreasing functions. }

\begin{equation}
\nonumber
\label{PositiveC}
    \Re\left[p^{*}\hat{C}_{ijkl}(p)\right]\geq 0
\end{equation}
Or, equivalently:
\begin{equation}
\label{PositiveC2}
    \Re\left[\frac{1}{p}\hat{C}_{ijkl}(p)\right]\geq 0
\end{equation}
Instead of setting $\phi(s)=e^{ps}$ one can also use $\phi(s)=e^{-iks}$ to have the Laplace transform of the stiffness matrix in the $k$-notation:
\begin{equation}
\label{PositiveC3}
    \Re\left[\frac{1}{-ik}\hat{C}_{ijkl}(k)\right]\geq 0
\end{equation}
In essence, we have shown that $\hat{\boldsymbol{C}}(p)/p$ is positive.
% It is notable to recall that both of equations \ref{PositiveC2} and \ref{PositiveC3} are pointing the real part of $\mathcal{L}\left[\int_{0}^{t}C_{ijkl}(v)dv\right]$ but using two different Laplace transform formulas. Moreover, from Eq. (\ref{PositiveC3}) one can conclude that the imaginary part of $i\mathcal{L}\left[\int_{0}^{t}C_{ijkl}(v)dv\right]$ (Laplace transform with respect to $k$) is non-negative definite which means that $-\frac{\hat{\boldsymbol{C}}(k)}{k}$ is a Herglotz tensor. 

\section{Acknowledgments}\label{sec:ack}
A.S. acknowledges support from the NSF CAREER grant \#1554033 to the Illinois Institute of Technology.


\begin{thebibliography}{54}%
\makeatletter
\providecommand \@ifxundefined [1]{%
 \@ifx{#1\undefined}
}%
\providecommand \@ifnum [1]{%
 \ifnum #1\expandafter \@firstoftwo
 \else \expandafter \@secondoftwo
 \fi
}%
\providecommand \@ifx [1]{%
 \ifx #1\expandafter \@firstoftwo
 \else \expandafter \@secondoftwo
 \fi
}%
\providecommand \natexlab [1]{#1}%
\providecommand \enquote  [1]{``#1''}%
\providecommand \bibnamefont  [1]{#1}%
\providecommand \bibfnamefont [1]{#1}%
\providecommand \citenamefont [1]{#1}%
\providecommand \href@noop [0]{\@secondoftwo}%
\providecommand \href [0]{\begingroup \@sanitize@url \@href}%
\providecommand \@href[1]{\@@startlink{#1}\@@href}%
\providecommand \@@href[1]{\endgroup#1\@@endlink}%
\providecommand \@sanitize@url [0]{\catcode `\\12\catcode `\$12\catcode
  `\&12\catcode `\#12\catcode `\^12\catcode `\_12\catcode `\%12\relax}%
\providecommand \@@startlink[1]{}%
\providecommand \@@endlink[0]{}%
\providecommand \url  [0]{\begingroup\@sanitize@url \@url }%
\providecommand \@url [1]{\endgroup\@href {#1}{\urlprefix }}%
\providecommand \urlprefix  [0]{URL }%
\providecommand \Eprint [0]{\href }%
\providecommand \doibase [0]{https://doi.org/}%
\providecommand \selectlanguage [0]{\@gobble}%
\providecommand \bibinfo  [0]{\@secondoftwo}%
\providecommand \bibfield  [0]{\@secondoftwo}%
\providecommand \translation [1]{[#1]}%
\providecommand \BibitemOpen [0]{}%
\providecommand \bibitemStop [0]{}%
\providecommand \bibitemNoStop [0]{.\EOS\space}%
\providecommand \EOS [0]{\spacefactor3000\relax}%
\providecommand \BibitemShut  [1]{\csname bibitem#1\endcsname}%
\let\auto@bib@innerbib\@empty
%</preamble>
\bibitem [{\citenamefont {Willis}(2009)}]{willis2009exact}%
  \BibitemOpen
  \bibfield  {author} {\bibinfo {author} {\bibfnamefont {J.~R.}\ \bibnamefont
  {Willis}},\ }\bibfield  {title} {\bibinfo {title} {{Exact effective relations
  for dynamics of a laminated body}},\ }\href@noop {} {\bibfield  {journal}
  {\bibinfo  {journal} {Mechanics of Materials}\ }\textbf {\bibinfo {volume}
  {41}},\ \bibinfo {pages} {385} (\bibinfo {year} {2009})}\BibitemShut
  {NoStop}%
\bibitem [{\citenamefont
  {Srivastava}(2015{\natexlab{a}})}]{Srivastava2015CausalityElastodynamics}%
  \BibitemOpen
  \bibfield  {author} {\bibinfo {author} {\bibfnamefont {A.}~\bibnamefont
  {Srivastava}},\ }\bibfield  {title} {\bibinfo {title} {{Causality and
  passivity in elastodynamics}},\ }\bibfield  {journal} {\bibinfo  {journal}
  {Proceedings of the Royal Society A: Mathematical, Physical and Engineering
  Sciences}\ }\textbf {\bibinfo {volume} {471}},\ \href
  {https://doi.org/10.1098/rspa.2015.0256} {10.1098/rspa.2015.0256} (\bibinfo
  {year} {2015}{\natexlab{a}})\BibitemShut {NoStop}%
\bibitem [{\citenamefont {Srivastava}(2020)}]{srivastava2020causality}%
  \BibitemOpen
  \bibfield  {author} {\bibinfo {author} {\bibfnamefont {A.}~\bibnamefont
  {Srivastava}},\ }\bibfield  {title} {\bibinfo {title} {Causality and
  passivity: From electromagnetism and network theory to metamaterials},\
  }\href@noop {} {\bibfield  {journal} {\bibinfo  {journal} {Mechanics of
  Materials}\ ,\ \bibinfo {pages} {103710}} (\bibinfo {year}
  {2020})}\BibitemShut {NoStop}%
\bibitem [{\citenamefont {Beltrami}\ and\ \citenamefont
  {Wohlers}(1966{\natexlab{a}})}]{beltrami1966distributionalb}%
  \BibitemOpen
  \bibfield  {author} {\bibinfo {author} {\bibfnamefont {E.~J.}\ \bibnamefont
  {Beltrami}}\ and\ \bibinfo {author} {\bibfnamefont {M.}~\bibnamefont
  {Wohlers}},\ }\bibfield  {title} {\bibinfo {title} {Distributional boundary
  values of functions holomorphic in a half plane},\ }\href@noop {} {\bibfield
  {journal} {\bibinfo  {journal} {Journal of Mathematics and Mechanics}\
  }\textbf {\bibinfo {volume} {15}},\ \bibinfo {pages} {137} (\bibinfo {year}
  {1966}{\natexlab{a}})}\BibitemShut {NoStop}%
\bibitem [{\citenamefont {Waters}(2000)}]{waters2000application}%
  \BibitemOpen
  \bibfield  {author} {\bibinfo {author} {\bibfnamefont {K.~R.}\ \bibnamefont
  {Waters}},\ }\emph {\bibinfo {title} {{On the application of the generalized
  Kramers-{Kronig} dispersion relations to ultrasonic propagation}}},\
  \href@noop {} {Ph.D. thesis},\ \bibinfo  {school} {Washington University}
  (\bibinfo {year} {2000})\BibitemShut {NoStop}%
\bibitem [{\citenamefont {Muhlestein}\ \emph {et~al.}(2016)\citenamefont
  {Muhlestein}, \citenamefont {Sieck}, \citenamefont {Al{\`u}},\ and\
  \citenamefont {Haberman}}]{muhlestein2016reciprocity}%
  \BibitemOpen
  \bibfield  {author} {\bibinfo {author} {\bibfnamefont {M.~B.}\ \bibnamefont
  {Muhlestein}}, \bibinfo {author} {\bibfnamefont {C.~F.}\ \bibnamefont
  {Sieck}}, \bibinfo {author} {\bibfnamefont {A.}~\bibnamefont {Al{\`u}}},\
  and\ \bibinfo {author} {\bibfnamefont {M.~R.}\ \bibnamefont {Haberman}},\
  }\bibfield  {title} {\bibinfo {title} {Reciprocity, passivity and causality
  in {Willis} materials},\ }\href@noop {} {\bibfield  {journal} {\bibinfo
  {journal} {Proceedings of the Royal Society A: Mathematical, Physical and
  Engineering Sciences}\ }\textbf {\bibinfo {volume} {472}},\ \bibinfo {pages}
  {20160604} (\bibinfo {year} {2016})}\BibitemShut {NoStop}%
\bibitem [{\citenamefont {Norris}(2018)}]{norris2018integral}%
  \BibitemOpen
  \bibfield  {author} {\bibinfo {author} {\bibfnamefont {A.~N.}\ \bibnamefont
  {Norris}},\ }\bibfield  {title} {\bibinfo {title} {Integral identities for
  reflection, transmission, and scattering coefficients},\ }\href@noop {}
  {\bibfield  {journal} {\bibinfo  {journal} {The Journal of the Acoustical
  Society of America}\ }\textbf {\bibinfo {volume} {144}},\ \bibinfo {pages}
  {2109} (\bibinfo {year} {2018})}\BibitemShut {NoStop}%
\bibitem [{\citenamefont {Nussenzveig}(1972)}]{nussenzveig1972causality}%
  \BibitemOpen
  \bibfield  {author} {\bibinfo {author} {\bibfnamefont {H.~M.}\ \bibnamefont
  {Nussenzveig}},\ }\href@noop {} {\emph {\bibinfo {title} {{Causality and
  dispersion relations}}}}\ (\bibinfo  {publisher} {Academic Press},\ \bibinfo
  {year} {1972})\BibitemShut {NoStop}%
\bibitem [{\citenamefont {Weaver}\ and\ \citenamefont
  {Pao}(1981)}]{weaver1981dispersion}%
  \BibitemOpen
  \bibfield  {author} {\bibinfo {author} {\bibfnamefont {R.~L.}\ \bibnamefont
  {Weaver}}\ and\ \bibinfo {author} {\bibfnamefont {Y.-H.}\ \bibnamefont
  {Pao}},\ }\bibfield  {title} {\bibinfo {title} {Dispersion relations for
  linear wave propagation in homogeneous and inhomogeneous media},\ }\href@noop
  {} {\bibfield  {journal} {\bibinfo  {journal} {Journal of Mathematical
  Physics}\ }\textbf {\bibinfo {volume} {22}},\ \bibinfo {pages} {1909}
  (\bibinfo {year} {1981})}\BibitemShut {NoStop}%
\bibitem [{\citenamefont {Ginzberg}(1955)}]{ginzberg1955}%
  \BibitemOpen
  \bibfield  {author} {\bibinfo {author} {\bibfnamefont {V.~L.}\ \bibnamefont
  {Ginzberg}},\ }\bibfield  {title} {\bibinfo {title} {Concerning the general
  relationship between absorption and dispersion of sound waves},\ }\href@noop
  {} {\bibfield  {journal} {\bibinfo  {journal} {Soviet Physical Acoustics}\
  }\textbf {\bibinfo {volume} {1}},\ \bibinfo {pages} {32} (\bibinfo {year}
  {1955})}\BibitemShut {NoStop}%
\bibitem [{\citenamefont {Futterman}(1962)}]{futterman1962dispersive}%
  \BibitemOpen
  \bibfield  {author} {\bibinfo {author} {\bibfnamefont {W.~I.}\ \bibnamefont
  {Futterman}},\ }\bibfield  {title} {\bibinfo {title} {Dispersive body
  waves},\ }\href@noop {} {\bibfield  {journal} {\bibinfo  {journal} {Journal
  of Geophysical research}\ }\textbf {\bibinfo {volume} {67}},\ \bibinfo
  {pages} {5279} (\bibinfo {year} {1962})}\BibitemShut {NoStop}%
\bibitem [{\citenamefont {Lamb~Jr}(1962)}]{lamb1962attenuation}%
  \BibitemOpen
  \bibfield  {author} {\bibinfo {author} {\bibfnamefont {G.~L.}\ \bibnamefont
  {Lamb~Jr}},\ }\bibfield  {title} {\bibinfo {title} {The attenuation of waves
  in a dispersive medium},\ }\href@noop {} {\bibfield  {journal} {\bibinfo
  {journal} {Journal of Geophysical Research}\ }\textbf {\bibinfo {volume}
  {67}},\ \bibinfo {pages} {5273} (\bibinfo {year} {1962})}\BibitemShut
  {NoStop}%
\bibitem [{\citenamefont {Strick}(1967)}]{strick1967determination}%
  \BibitemOpen
  \bibfield  {author} {\bibinfo {author} {\bibfnamefont {E.}~\bibnamefont
  {Strick}},\ }\bibfield  {title} {\bibinfo {title} {The determination of q,
  dynamic viscosity and transient creep curves from wave propagation
  measurements},\ }\href@noop {} {\bibfield  {journal} {\bibinfo  {journal}
  {Geophysical Journal International}\ }\textbf {\bibinfo {volume} {13}},\
  \bibinfo {pages} {197} (\bibinfo {year} {1967})}\BibitemShut {NoStop}%
\bibitem [{\citenamefont {Azimi}(1968)}]{azimi1968impulse}%
  \BibitemOpen
  \bibfield  {author} {\bibinfo {author} {\bibfnamefont {S.~A.}\ \bibnamefont
  {Azimi}},\ }\bibfield  {title} {\bibinfo {title} {Impulse and transient
  characteristics of media with linear and quadratic absorption laws,
  {Izvestiya}},\ }\href@noop {} {\bibfield  {journal} {\bibinfo  {journal}
  {Physics of the Solid Earth}\ ,\ \bibinfo {pages} {88}} (\bibinfo {year}
  {1968})}\BibitemShut {NoStop}%
\bibitem [{\citenamefont {Randall}(1976)}]{randall1976attenuative}%
  \BibitemOpen
  \bibfield  {author} {\bibinfo {author} {\bibfnamefont {M.~J.}\ \bibnamefont
  {Randall}},\ }\bibfield  {title} {\bibinfo {title} {Attenuative dispersion
  and frequency shifts of the {Earth}'s free oscillations},\ }\href@noop {}
  {\bibfield  {journal} {\bibinfo  {journal} {Physics of the Earth and
  Planetary Interiors}\ }\textbf {\bibinfo {volume} {12}},\ \bibinfo {pages}
  {P1} (\bibinfo {year} {1976})}\BibitemShut {NoStop}%
\bibitem [{\citenamefont {Liu}\ \emph {et~al.}(1976)\citenamefont {Liu},
  \citenamefont {Anderson},\ and\ \citenamefont {Kanamori}}]{liu1976velocity}%
  \BibitemOpen
  \bibfield  {author} {\bibinfo {author} {\bibfnamefont {H.-P.}\ \bibnamefont
  {Liu}}, \bibinfo {author} {\bibfnamefont {D.~L.}\ \bibnamefont {Anderson}},\
  and\ \bibinfo {author} {\bibfnamefont {H.}~\bibnamefont {Kanamori}},\
  }\bibfield  {title} {\bibinfo {title} {Velocity dispersion due to
  anelasticity; implications for seismology and mantle composition},\
  }\href@noop {} {\bibfield  {journal} {\bibinfo  {journal} {Geophysical
  Journal International}\ }\textbf {\bibinfo {volume} {47}},\ \bibinfo {pages}
  {41} (\bibinfo {year} {1976})}\BibitemShut {NoStop}%
\bibitem [{\citenamefont {Hamilton}(1970)}]{hamilton1970sound}%
  \BibitemOpen
  \bibfield  {author} {\bibinfo {author} {\bibfnamefont {E.~L.}\ \bibnamefont
  {Hamilton}},\ }\bibfield  {title} {\bibinfo {title} {Sound velocity and
  related properties of marine sediments, north pacific},\ }\href@noop {}
  {\bibfield  {journal} {\bibinfo  {journal} {Journal of Geophysical Research}\
  }\textbf {\bibinfo {volume} {75}},\ \bibinfo {pages} {4423} (\bibinfo {year}
  {1970})}\BibitemShut {NoStop}%
\bibitem [{\citenamefont {Horton~Sr}(1974)}]{horton1974dispersion}%
  \BibitemOpen
  \bibfield  {author} {\bibinfo {author} {\bibfnamefont {C.}~\bibnamefont
  {Horton~Sr}},\ }\bibfield  {title} {\bibinfo {title} {Dispersion
  relationships in sediments and sea water},\ }\href@noop {} {\bibfield
  {journal} {\bibinfo  {journal} {The Journal of the Acoustical Society of
  America}\ }\textbf {\bibinfo {volume} {55}},\ \bibinfo {pages} {547}
  (\bibinfo {year} {1974})}\BibitemShut {NoStop}%
\bibitem [{\citenamefont {Horton~Sr}(1981)}]{horton1981comment}%
  \BibitemOpen
  \bibfield  {author} {\bibinfo {author} {\bibfnamefont {C.}~\bibnamefont
  {Horton~Sr}},\ }\bibfield  {title} {\bibinfo {title} {Comment on
  {Kramers-Kronig} relationship between ultrasonic attenuation and phase
  velocity},\ }\href@noop {} {\bibfield  {journal} {\bibinfo  {journal}
  {Journal of the Acoustical Society of America}\ }\textbf {\bibinfo {volume}
  {70}} (\bibinfo {year} {1981})}\BibitemShut {NoStop}%
\bibitem [{\citenamefont {Waters}\ \emph {et~al.}(1999)\citenamefont {Waters},
  \citenamefont {Hughes}, \citenamefont {Mobley}, \citenamefont
  {Brandenburger},\ and\ \citenamefont {Miller}}]{waters1999kramers}%
  \BibitemOpen
  \bibfield  {author} {\bibinfo {author} {\bibfnamefont {K.~R.}\ \bibnamefont
  {Waters}}, \bibinfo {author} {\bibfnamefont {M.~S.}\ \bibnamefont {Hughes}},
  \bibinfo {author} {\bibfnamefont {J.}~\bibnamefont {Mobley}}, \bibinfo
  {author} {\bibfnamefont {G.~H.}\ \bibnamefont {Brandenburger}},\ and\
  \bibinfo {author} {\bibfnamefont {J.~G.}\ \bibnamefont {Miller}},\ }\bibfield
   {title} {\bibinfo {title} {Kramers-{Kronig} dispersion relations for
  ultrasonic attenuation obeying a frequency power law},\ }in\ \href@noop {}
  {\emph {\bibinfo {booktitle} {1999 IEEE Ultrasonics Symposium. Proceedings.
  International Symposium (Cat. No. 99CH37027)}}},\ Vol.~\bibinfo {volume} {1}\
  (\bibinfo {organization} {IEEE},\ \bibinfo {year} {1999})\ pp.\ \bibinfo
  {pages} {537--541}\BibitemShut {NoStop}%
\bibitem [{\citenamefont {Waters}\ \emph {et~al.}(2003)\citenamefont {Waters},
  \citenamefont {Hughes}, \citenamefont {Mobley},\ and\ \citenamefont
  {Miller}}]{waters2003differential}%
  \BibitemOpen
  \bibfield  {author} {\bibinfo {author} {\bibfnamefont {K.~R.}\ \bibnamefont
  {Waters}}, \bibinfo {author} {\bibfnamefont {M.~S.}\ \bibnamefont {Hughes}},
  \bibinfo {author} {\bibfnamefont {J.}~\bibnamefont {Mobley}},\ and\ \bibinfo
  {author} {\bibfnamefont {J.~G.}\ \bibnamefont {Miller}},\ }\bibfield  {title}
  {\bibinfo {title} {Differential forms of the {Kramers-Kronig} dispersion
  relations},\ }\href@noop {} {\bibfield  {journal} {\bibinfo  {journal} {IEEE
  transactions on ultrasonics, ferroelectrics, and frequency control}\ }\textbf
  {\bibinfo {volume} {50}},\ \bibinfo {pages} {68} (\bibinfo {year}
  {2003})}\BibitemShut {NoStop}%
\bibitem [{\citenamefont {Waters}\ \emph {et~al.}(2005)\citenamefont {Waters},
  \citenamefont {Mobley},\ and\ \citenamefont {Miller}}]{waters2005causality}%
  \BibitemOpen
  \bibfield  {author} {\bibinfo {author} {\bibfnamefont {K.~R.}\ \bibnamefont
  {Waters}}, \bibinfo {author} {\bibfnamefont {J.}~\bibnamefont {Mobley}},\
  and\ \bibinfo {author} {\bibfnamefont {J.~G.}\ \bibnamefont {Miller}},\
  }\bibfield  {title} {\bibinfo {title} {{Causality-imposed ({Kramers-Kronig})
  relationships between attenuation and dispersion}},\ }\href@noop {}
  {\bibfield  {journal} {\bibinfo  {journal} {Ultrasonics, Ferroelectrics, and
  Frequency Control, IEEE Transactions on}\ }\textbf {\bibinfo {volume} {52}},\
  \bibinfo {pages} {822} (\bibinfo {year} {2005})}\BibitemShut {NoStop}%
\bibitem [{\citenamefont {Herglotz}(1911{\natexlab{a}})}]{herglotz1911uber}%
  \BibitemOpen
  \bibfield  {author} {\bibinfo {author} {\bibfnamefont {G.}~\bibnamefont
  {Herglotz}},\ }\bibfield  {title} {\bibinfo {title} {{\"{U}ber Potenzreihen
  mit positivem, reelen Teil im Einheitskreis}},\ }\href@noop {} {\bibfield
  {journal} {\bibinfo  {journal} {Ber. Verhandl. Sachs Akad. Wiss. Leipzig,
  Math.-Phys. Kl.}\ }\textbf {\bibinfo {volume} {63}},\ \bibinfo {pages} {501}
  (\bibinfo {year} {1911}{\natexlab{a}})}\BibitemShut {NoStop}%
\bibitem [{\citenamefont {Youla}(1958)}]{youla1958representation}%
  \BibitemOpen
  \bibfield  {author} {\bibinfo {author} {\bibfnamefont {D.}~\bibnamefont
  {Youla}},\ }\bibfield  {title} {\bibinfo {title} {Representation theory of
  linear passive networks},\ }in\ \href@noop {} {\emph {\bibinfo {booktitle}
  {MRI Report No. R-655-58}}}\ (\bibinfo  {publisher} {Poly. Inst. of Bklyn},\
  \bibinfo {year} {1958})\BibitemShut {NoStop}%
\bibitem [{\citenamefont {Beltrami}(1967)}]{beltrami1967linear}%
  \BibitemOpen
  \bibfield  {author} {\bibinfo {author} {\bibfnamefont {E.~J.}\ \bibnamefont
  {Beltrami}},\ }\bibfield  {title} {\bibinfo {title} {Linear dissipative
  systems, nonnegative definite distributional kernels, and the boundary values
  of bounded-real and positive-real matrices},\ }\href@noop {} {\bibfield
  {journal} {\bibinfo  {journal} {Journal of Mathematical Analysis and
  Applications}\ }\textbf {\bibinfo {volume} {19}},\ \bibinfo {pages} {231}
  (\bibinfo {year} {1967})}\BibitemShut {NoStop}%
\bibitem [{\citenamefont
  {Srivastava}(2015{\natexlab{b}})}]{srivastava2015elastic}%
  \BibitemOpen
  \bibfield  {author} {\bibinfo {author} {\bibfnamefont {A.}~\bibnamefont
  {Srivastava}},\ }\bibfield  {title} {\bibinfo {title} {{Elastic metamaterials
  and dynamic homogenization: a review}},\ }\href@noop {} {\bibfield  {journal}
  {\bibinfo  {journal} {International Journal of Smart and Nano Materials}\
  }\textbf {\bibinfo {volume} {6}},\ \bibinfo {pages} {41} (\bibinfo {year}
  {2015}{\natexlab{b}})}\BibitemShut {NoStop}%
\bibitem [{\citenamefont
  {Zemanian}(1965{\natexlab{a}})}]{zemanian1965distribution}%
  \BibitemOpen
  \bibfield  {author} {\bibinfo {author} {\bibfnamefont {A.~H.}\ \bibnamefont
  {Zemanian}},\ }\href@noop {} {\emph {\bibinfo {title} {{Distribution theory
  and transform analysis: an introduction to generalized functions, with
  applications}}}}\ (\bibinfo  {publisher} {Courier Corporation},\ \bibinfo
  {year} {1965})\BibitemShut {NoStop}%
\bibitem [{\citenamefont {Nedic}(2017)}]{Herglotzthesis}%
  \BibitemOpen
  \bibfield  {author} {\bibinfo {author} {\bibfnamefont {M.}~\bibnamefont
  {Nedic}},\ }\bibfield  {title} {\bibinfo {title} {Integral representations of
  herglotz-nevanlinna functions},\ }\href
  {http://urn.kb.se/resolve?urn=urn%3Anbn%3Ase%3Asu%3Adiva-138962} {\bibfield
  {journal} {\bibinfo  {journal} {Stockholm, Department of Mathematics}\ ,\
  \bibinfo {pages} {118}} (\bibinfo {year} {2017})}\BibitemShut {NoStop}%
\bibitem [{\citenamefont {Bernland}\ \emph {et~al.}(2011)\citenamefont
  {Bernland}, \citenamefont {Luger},\ and\ \citenamefont
  {Gustafsson}}]{sumrules}%
  \BibitemOpen
  \bibfield  {author} {\bibinfo {author} {\bibfnamefont {A.}~\bibnamefont
  {Bernland}}, \bibinfo {author} {\bibfnamefont {A.}~\bibnamefont {Luger}},\
  and\ \bibinfo {author} {\bibfnamefont {M.}~\bibnamefont {Gustafsson}},\
  }\bibfield  {title} {\bibinfo {title} {Sum rules and constraints on passive
  systems},\ }\href {https://doi.org/10.1088/1751-8113/44/14/145205} {\bibfield
   {journal} {\bibinfo  {journal} {Journal of Physics A: Mathematical and
  Theoretical}\ }\textbf {\bibinfo {volume} {44}},\ \bibinfo {pages} {145205}
  (\bibinfo {year} {2011})}\BibitemShut {NoStop}%
\bibitem [{\citenamefont {Herglotz}(1911{\natexlab{b}})}]{herglotzorig}%
  \BibitemOpen
  \bibfield  {author} {\bibinfo {author} {\bibfnamefont {G.}~\bibnamefont
  {Herglotz}},\ }\bibfield  {title} {\bibinfo {title} {Uber potenzreihen mit
  positivem, reelen teil im einheitskreis},\ }\href@noop {} {\bibfield
  {journal} {\bibinfo  {journal} {Ber. Verhandl. Sachs Akad. Wiss. Leipzig,
  Math.-Phys. Kl.}\ }\textbf {\bibinfo {volume} {63}},\ \bibinfo {pages} {501}
  (\bibinfo {year} {1911}{\natexlab{b}})}\BibitemShut {NoStop}%
\bibitem [{\citenamefont {Zemanian}(1963)}]{zemanian1963n}%
  \BibitemOpen
  \bibfield  {author} {\bibinfo {author} {\bibfnamefont {A.~H.}\ \bibnamefont
  {Zemanian}},\ }\bibfield  {title} {\bibinfo {title} {{An N-port realizability
  theory based on the theory of distributions}},\ }\href@noop {} {\bibfield
  {journal} {\bibinfo  {journal} {Circuit Theory, IEEE Transactions on}\
  }\textbf {\bibinfo {volume} {10}},\ \bibinfo {pages} {265} (\bibinfo {year}
  {1963})}\BibitemShut {NoStop}%
\bibitem [{\citenamefont {Meixner}(1959)}]{meixnernetwork}%
  \BibitemOpen
  \bibfield  {author} {\bibinfo {author} {\bibfnamefont {J.}~\bibnamefont
  {Meixner}},\ }\bibfield  {title} {\bibinfo {title} {Network theory and its
  relation to the theory of linear systems},\ }\href@noop {} {\bibfield
  {journal} {\bibinfo  {journal} {IRE Transactions on Antennas and
  Propagation}\ }\textbf {\bibinfo {volume} {7}},\ \bibinfo {pages} {435}
  (\bibinfo {year} {1959})}\BibitemShut {NoStop}%
\bibitem [{\citenamefont {Zemanian}(1965{\natexlab{b}})}]{CharacterZemanian}%
  \BibitemOpen
  \bibfield  {author} {\bibinfo {author} {\bibfnamefont {A.~H.}\ \bibnamefont
  {Zemanian}},\ }\bibfield  {title} {\bibinfo {title} {A characterization of
  the inverse laplace transforms of rational positive-real matrices},\ }\href
  {https://doi.org/10.1137/0113028} {\bibfield  {journal} {\bibinfo  {journal}
  {Journal of the Society for Industrial and Applied Mathematics}\ }\textbf
  {\bibinfo {volume} {13}},\ \bibinfo {pages} {463} (\bibinfo {year}
  {1965}{\natexlab{b}})},\ \Eprint
  {https://arxiv.org/abs/https://doi.org/10.1137/0113028}
  {https://doi.org/10.1137/0113028} \BibitemShut {NoStop}%
\bibitem [{\citenamefont {Konig}\ and\ \citenamefont
  {Zemanian}(1965)}]{Necandsuf}%
  \BibitemOpen
  \bibfield  {author} {\bibinfo {author} {\bibfnamefont {H.}~\bibnamefont
  {Konig}}\ and\ \bibinfo {author} {\bibfnamefont {A.~H.}\ \bibnamefont
  {Zemanian}},\ }\bibfield  {title} {\bibinfo {title} {Necessary and sufficient
  conditions for a matrix distribution to have a positive-real laplace
  transform},\ }\href {http://www.jstor.org/stable/2946423} {\bibfield
  {journal} {\bibinfo  {journal} {Journal of the Society for Industrial and
  Applied Mathematics}\ }\textbf {\bibinfo {volume} {13}},\ \bibinfo {pages}
  {1036} (\bibinfo {year} {1965})}\BibitemShut {NoStop}%
\bibitem [{\citenamefont {Cassier}\ and\ \citenamefont
  {Milton}(2017)}]{cassier2017bounds}%
  \BibitemOpen
  \bibfield  {author} {\bibinfo {author} {\bibfnamefont {M.}~\bibnamefont
  {Cassier}}\ and\ \bibinfo {author} {\bibfnamefont {G.~W.}\ \bibnamefont
  {Milton}},\ }\bibfield  {title} {\bibinfo {title} {Bounds on {Herglotz}
  functions and fundamental limits of broadband passive quasistatic cloaking},\
  }\href@noop {} {\bibfield  {journal} {\bibinfo  {journal} {Journal of
  Mathematical Physics}\ }\textbf {\bibinfo {volume} {58}},\ \bibinfo {pages}
  {071504} (\bibinfo {year} {2017})}\BibitemShut {NoStop}%
\bibitem [{\citenamefont {Cauer}(1932)}]{cauer1932poisson}%
  \BibitemOpen
  \bibfield  {author} {\bibinfo {author} {\bibfnamefont {W.}~\bibnamefont
  {Cauer}},\ }\bibfield  {title} {\bibinfo {title} {{The Poisson integral for
  functions with positive real part}},\ }\href@noop {} {\bibfield  {journal}
  {\bibinfo  {journal} {Bulletin of the American Mathematical Society}\
  }\textbf {\bibinfo {volume} {38}},\ \bibinfo {pages} {713} (\bibinfo {year}
  {1932})}\BibitemShut {NoStop}%
\bibitem [{\citenamefont {Phillips}(1950)}]{phillips1950fourier}%
  \BibitemOpen
  \bibfield  {author} {\bibinfo {author} {\bibfnamefont {R.}~\bibnamefont
  {Phillips}},\ }\bibfield  {title} {\bibinfo {title} {{On Fourier-Stieltjes
  integrals}},\ }\href@noop {} {\bibfield  {journal} {\bibinfo  {journal}
  {Transactions of the American Mathematical Society}\ }\textbf {\bibinfo
  {volume} {69}},\ \bibinfo {pages} {312} (\bibinfo {year} {1950})}\BibitemShut
  {NoStop}%
\bibitem [{\citenamefont {Beltrami}\ and\ \citenamefont
  {Wohlers}(1966{\natexlab{b}})}]{beltrami1966}%
  \BibitemOpen
  \bibfield  {author} {\bibinfo {author} {\bibfnamefont {E.~J.}\ \bibnamefont
  {Beltrami}}\ and\ \bibinfo {author} {\bibfnamefont {M.}~\bibnamefont
  {Wohlers}},\ }\href@noop {} {\emph {\bibinfo {title} {{Distributions and the
  Boundary Values of Analytic Functions}}}}\ (\bibinfo  {publisher} {Academic
  Press},\ \bibinfo {year} {1966})\BibitemShut {NoStop}%
\bibitem [{\citenamefont
  {Srivastava}(2015{\natexlab{c}})}]{srivastava2015causality}%
  \BibitemOpen
  \bibfield  {author} {\bibinfo {author} {\bibfnamefont {A.}~\bibnamefont
  {Srivastava}},\ }\bibfield  {title} {\bibinfo {title} {{Causality and
  passivity in elastodynamics}},\ }in\ \href@noop {} {\emph {\bibinfo
  {booktitle} {Proc. R. Soc. A}}},\ Vol.\ \bibinfo {volume} {471}\ (\bibinfo
  {organization} {The Royal Society},\ \bibinfo {year} {2015})\ p.\ \bibinfo
  {pages} {20150256}\BibitemShut {NoStop}%
\bibitem [{\citenamefont {McMillan}(1952)}]{mcmillan1952introduction}%
  \BibitemOpen
  \bibfield  {author} {\bibinfo {author} {\bibfnamefont {B.}~\bibnamefont
  {McMillan}},\ }\bibfield  {title} {\bibinfo {title} {Introduction to formal
  realizability theory—i},\ }\href@noop {} {\bibfield  {journal} {\bibinfo
  {journal} {Bell System Technical Journal}\ }\textbf {\bibinfo {volume}
  {31}},\ \bibinfo {pages} {217} (\bibinfo {year} {1952})}\BibitemShut
  {NoStop}%
\bibitem [{\citenamefont {Achenbach}(1984)}]{achenbach1984wave}%
  \BibitemOpen
  \bibfield  {author} {\bibinfo {author} {\bibfnamefont {J.}~\bibnamefont
  {Achenbach}},\ }\href@noop {} {\emph {\bibinfo {title} {{Wave propagation in
  elastic solids}}}}\ (\bibinfo  {publisher} {Elsevier},\ \bibinfo {year}
  {1984})\BibitemShut {NoStop}%
\bibitem [{\citenamefont {Willis}(1997)}]{willis1997dynamics}%
  \BibitemOpen
  \bibfield  {author} {\bibinfo {author} {\bibfnamefont {J.~R.}\ \bibnamefont
  {Willis}},\ }\bibfield  {title} {\bibinfo {title} {{Dynamics of
  composites}},\ }in\ \href@noop {} {\emph {\bibinfo {booktitle} {Continuum
  micromechanics}}}\ (\bibinfo {organization} {Springer-Verlag New York,
  Inc.},\ \bibinfo {year} {1997})\ pp.\ \bibinfo {pages} {265--290}\BibitemShut
  {NoStop}%
\bibitem [{\citenamefont {Pernas-Salom\'on}\ and\ \citenamefont
  {Shmuel}(2020)}]{GalforWillis}%
  \BibitemOpen
  \bibfield  {author} {\bibinfo {author} {\bibfnamefont {R.}~\bibnamefont
  {Pernas-Salom\'on}}\ and\ \bibinfo {author} {\bibfnamefont {G.}~\bibnamefont
  {Shmuel}},\ }\bibfield  {title} {\bibinfo {title} {{Fundamental Principles
  for Generalized Willis Metamaterials}},\ }\href
  {https://doi.org/10.1103/PhysRevApplied.14.064005} {\bibfield  {journal}
  {\bibinfo  {journal} {Phys. Rev. Applied}\ }\textbf {\bibinfo {volume}
  {14}},\ \bibinfo {pages} {064005} (\bibinfo {year} {2020})}\BibitemShut
  {NoStop}%
\bibitem [{\citenamefont {Nemat-Nasser}\ \emph {et~al.}(2011)\citenamefont
  {Nemat-Nasser}, \citenamefont {Willis}, \citenamefont {Srivastava},\ and\
  \citenamefont {Amirkhizi}}]{nemat2011homogenization}%
  \BibitemOpen
  \bibfield  {author} {\bibinfo {author} {\bibfnamefont {S.}~\bibnamefont
  {Nemat-Nasser}}, \bibinfo {author} {\bibfnamefont {J.~R.}\ \bibnamefont
  {Willis}}, \bibinfo {author} {\bibfnamefont {A.}~\bibnamefont {Srivastava}},\
  and\ \bibinfo {author} {\bibfnamefont {A.~V.}\ \bibnamefont {Amirkhizi}},\
  }\bibfield  {title} {\bibinfo {title} {{Homogenization of periodic elastic
  composites and locally resonant sonic materials}},\ }\href@noop {} {\bibfield
   {journal} {\bibinfo  {journal} {Physical Review B}\ }\textbf {\bibinfo
  {volume} {83}},\ \bibinfo {pages} {104103} (\bibinfo {year}
  {2011})}\BibitemShut {NoStop}%
\bibitem [{\citenamefont {Alizadeh}\ and\ \citenamefont
  {Amirkhizi}(2021)}]{alizadeh2021overall}%
  \BibitemOpen
  \bibfield  {author} {\bibinfo {author} {\bibfnamefont {V.}~\bibnamefont
  {Alizadeh}}\ and\ \bibinfo {author} {\bibfnamefont {A.~V.}\ \bibnamefont
  {Amirkhizi}},\ }\bibfield  {title} {\bibinfo {title} {{Overall dynamic
  properties of locally resonant viscoelastic layered media based on consistent
  field integration for oblique anti-plane shear waves}},\ }\href@noop {}
  {\bibfield  {journal} {\bibinfo  {journal} {arXiv preprint arXiv:2104.10571}\
  } (\bibinfo {year} {2021})}\BibitemShut {NoStop}%
\bibitem [{\citenamefont {Aghighi}\ \emph {et~al.}(2019)\citenamefont
  {Aghighi}, \citenamefont {Morris},\ and\ \citenamefont
  {Amirkhizi}}]{aghighi2019low}%
  \BibitemOpen
  \bibfield  {author} {\bibinfo {author} {\bibfnamefont {F.}~\bibnamefont
  {Aghighi}}, \bibinfo {author} {\bibfnamefont {J.}~\bibnamefont {Morris}},\
  and\ \bibinfo {author} {\bibfnamefont {A.~V.}\ \bibnamefont {Amirkhizi}},\
  }\bibfield  {title} {\bibinfo {title} {{Low-frequency micro-structured
  mechanical metamaterials}},\ }\href@noop {} {\bibfield  {journal} {\bibinfo
  {journal} {Mechanics of Materials}\ }\textbf {\bibinfo {volume} {130}},\
  \bibinfo {pages} {65} (\bibinfo {year} {2019})}\BibitemShut {NoStop}%
\bibitem [{\citenamefont {Amirkhizi}\ and\ \citenamefont
  {Alizadeh}(2018)}]{amirkhizi2018overall}%
  \BibitemOpen
  \bibfield  {author} {\bibinfo {author} {\bibfnamefont {A.~V.}\ \bibnamefont
  {Amirkhizi}}\ and\ \bibinfo {author} {\bibfnamefont {V.}~\bibnamefont
  {Alizadeh}},\ }\bibfield  {title} {\bibinfo {title} {{Overall constitutive
  description of symmetric layered media based on scattering of oblique SH
  waves}},\ }\href@noop {} {\bibfield  {journal} {\bibinfo  {journal} {Wave
  Motion}\ }\textbf {\bibinfo {volume} {83}},\ \bibinfo {pages} {214} (\bibinfo
  {year} {2018})}\BibitemShut {NoStop}%
\bibitem [{\citenamefont {Shmuel}\ \emph {et~al.}(2021)\citenamefont {Shmuel},
  \citenamefont {Pernas-Salomón}, \citenamefont {Muhafra}, \citenamefont
  {Kosta}, \citenamefont {Torrent}, \citenamefont {Haberman},\ and\
  \citenamefont {Norris}}]{Norris1}%
  \BibitemOpen
  \bibfield  {author} {\bibinfo {author} {\bibfnamefont {G.}~\bibnamefont
  {Shmuel}}, \bibinfo {author} {\bibfnamefont {R.}~\bibnamefont
  {Pernas-Salomón}}, \bibinfo {author} {\bibfnamefont {A.}~\bibnamefont
  {Muhafra}}, \bibinfo {author} {\bibfnamefont {M.}~\bibnamefont {Kosta}},
  \bibinfo {author} {\bibfnamefont {D.}~\bibnamefont {Torrent}}, \bibinfo
  {author} {\bibfnamefont {M.~R.}\ \bibnamefont {Haberman}},\ and\ \bibinfo
  {author} {\bibfnamefont {A.~N.}\ \bibnamefont {Norris}},\ }\bibfield  {title}
  {\bibinfo {title} {{The electromomentum coupling in generalized Willis
  media}},\ }\href {https://doi.org/10.1121/10.0004412} {\bibfield  {journal}
  {\bibinfo  {journal} {The Journal of the Acoustical Society of America}\
  }\textbf {\bibinfo {volume} {149}},\ \bibinfo {pages} {A23} (\bibinfo {year}
  {2021})},\ \Eprint {https://arxiv.org/abs/https://doi.org/10.1121/10.0004412}
  {https://doi.org/10.1121/10.0004412} \BibitemShut {NoStop}%
\bibitem [{\citenamefont {Nassar}\ \emph {et~al.}(2020)\citenamefont {Nassar},
  \citenamefont {Yousefzadeh}, \citenamefont {Fleury}, \citenamefont {Ruzzene},
  \citenamefont {Al{\`u}}, \citenamefont {Daraio}, \citenamefont {Norris},
  \citenamefont {Huang},\ and\ \citenamefont
  {Haberman}}]{nassar2020nonreciprocity}%
  \BibitemOpen
  \bibfield  {author} {\bibinfo {author} {\bibfnamefont {H.}~\bibnamefont
  {Nassar}}, \bibinfo {author} {\bibfnamefont {B.}~\bibnamefont {Yousefzadeh}},
  \bibinfo {author} {\bibfnamefont {R.}~\bibnamefont {Fleury}}, \bibinfo
  {author} {\bibfnamefont {M.}~\bibnamefont {Ruzzene}}, \bibinfo {author}
  {\bibfnamefont {A.}~\bibnamefont {Al{\`u}}}, \bibinfo {author} {\bibfnamefont
  {C.}~\bibnamefont {Daraio}}, \bibinfo {author} {\bibfnamefont {A.~N.}\
  \bibnamefont {Norris}}, \bibinfo {author} {\bibfnamefont {G.}~\bibnamefont
  {Huang}},\ and\ \bibinfo {author} {\bibfnamefont {M.~R.}\ \bibnamefont
  {Haberman}},\ }\bibfield  {title} {\bibinfo {title} {{Nonreciprocity in
  acoustic and elastic materials}},\ }\href@noop {} {\bibfield  {journal}
  {\bibinfo  {journal} {Nature Reviews Materials}\ }\textbf {\bibinfo {volume}
  {5}},\ \bibinfo {pages} {667} (\bibinfo {year} {2020})}\BibitemShut {NoStop}%
\bibitem [{\citenamefont {Nassar}\ \emph {et~al.}(2017)\citenamefont {Nassar},
  \citenamefont {Xu}, \citenamefont {Norris},\ and\ \citenamefont
  {Huang}}]{nassar2017modulated}%
  \BibitemOpen
  \bibfield  {author} {\bibinfo {author} {\bibfnamefont {H.}~\bibnamefont
  {Nassar}}, \bibinfo {author} {\bibfnamefont {X.~C.}\ \bibnamefont {Xu}},
  \bibinfo {author} {\bibfnamefont {A.~N.}\ \bibnamefont {Norris}},\ and\
  \bibinfo {author} {\bibfnamefont {G.~L.}\ \bibnamefont {Huang}},\ }\bibfield
  {title} {\bibinfo {title} {{Modulated phononic crystals: Non-reciprocal wave
  propagation and Willis materials}},\ }\href@noop {} {\bibfield  {journal}
  {\bibinfo  {journal} {Journal of the Mechanics and Physics of Solids}\
  }\textbf {\bibinfo {volume} {101}},\ \bibinfo {pages} {10} (\bibinfo {year}
  {2017})}\BibitemShut {NoStop}%
\bibitem [{\citenamefont {Chen}\ \emph {et~al.}(2020)\citenamefont {Chen},
  \citenamefont {Li}, \citenamefont {Hu}, \citenamefont {Haberman},\ and\
  \citenamefont {Huang}}]{chen2020active}%
  \BibitemOpen
  \bibfield  {author} {\bibinfo {author} {\bibfnamefont {Y.}~\bibnamefont
  {Chen}}, \bibinfo {author} {\bibfnamefont {X.}~\bibnamefont {Li}}, \bibinfo
  {author} {\bibfnamefont {G.}~\bibnamefont {Hu}}, \bibinfo {author}
  {\bibfnamefont {M.~R.}\ \bibnamefont {Haberman}},\ and\ \bibinfo {author}
  {\bibfnamefont {G.}~\bibnamefont {Huang}},\ }\bibfield  {title} {\bibinfo
  {title} {{An active mechanical Willis meta-layer with asymmetric
  polarizabilities}},\ }\href@noop {} {\bibfield  {journal} {\bibinfo
  {journal} {Nature communications}\ }\textbf {\bibinfo {volume} {11}},\
  \bibinfo {pages} {1} (\bibinfo {year} {2020})}\BibitemShut {NoStop}%
\bibitem [{\citenamefont {Muhlestein}\ \emph {et~al.}(2017)\citenamefont
  {Muhlestein}, \citenamefont {Sieck}, \citenamefont {Wilson},\ and\
  \citenamefont {Haberman}}]{muhlestein2017experimental}%
  \BibitemOpen
  \bibfield  {author} {\bibinfo {author} {\bibfnamefont {M.~B.}\ \bibnamefont
  {Muhlestein}}, \bibinfo {author} {\bibfnamefont {C.~F.}\ \bibnamefont
  {Sieck}}, \bibinfo {author} {\bibfnamefont {P.~S.}\ \bibnamefont {Wilson}},\
  and\ \bibinfo {author} {\bibfnamefont {M.~R.}\ \bibnamefont {Haberman}},\
  }\bibfield  {title} {\bibinfo {title} {{Experimental evidence of Willis
  coupling in a one-dimensional effective material element}},\ }\href@noop {}
  {\bibfield  {journal} {\bibinfo  {journal} {Nature communications}\ }\textbf
  {\bibinfo {volume} {8}},\ \bibinfo {pages} {1} (\bibinfo {year}
  {2017})}\BibitemShut {NoStop}%
\bibitem [{\citenamefont {Muhlestein}\ and\ \citenamefont
  {Haberman}(2017)}]{muhlestein2017analysis}%
  \BibitemOpen
  \bibfield  {author} {\bibinfo {author} {\bibfnamefont {M.}~\bibnamefont
  {Muhlestein}}\ and\ \bibinfo {author} {\bibfnamefont {M.~R.}\ \bibnamefont
  {Haberman}},\ }\bibfield  {title} {\bibinfo {title} {{Analysis of
  one-dimensional wave phenomena in Willis materials}},\ }in\ \href@noop {}
  {\emph {\bibinfo {booktitle} {Proceedings of Meetings on Acoustics
  173EAA}}},\ Vol.~\bibinfo {volume} {30}\ (\bibinfo {organization} {Acoustical
  Society of America},\ \bibinfo {year} {2017})\ p.\ \bibinfo {pages}
  {065017}\BibitemShut {NoStop}%
\bibitem [{\citenamefont {Sieck}\ \emph {et~al.}(2017)\citenamefont {Sieck},
  \citenamefont {Al\`u},\ and\ \citenamefont {Haberman}}]{PhysRevB.96.104303}%
  \BibitemOpen
  \bibfield  {author} {\bibinfo {author} {\bibfnamefont {C.~F.}\ \bibnamefont
  {Sieck}}, \bibinfo {author} {\bibfnamefont {A.}~\bibnamefont {Al\`u}},\ and\
  \bibinfo {author} {\bibfnamefont {M.~R.}\ \bibnamefont {Haberman}},\
  }\bibfield  {title} {\bibinfo {title} {{Origins of Willis coupling and
  acoustic bianisotropy in acoustic metamaterials through source-driven
  homogenization}},\ }\href {https://doi.org/10.1103/PhysRevB.96.104303}
  {\bibfield  {journal} {\bibinfo  {journal} {Phys. Rev. B}\ }\textbf {\bibinfo
  {volume} {96}},\ \bibinfo {pages} {104303} (\bibinfo {year}
  {2017})}\BibitemShut {NoStop}%
\end{thebibliography}
\end{document}